%% file: main.tex
\documentclass[twocolumn]{aastex63}
\pdfoutput=1 
\usepackage{amsmath,amstext}
\usepackage[T1]{fontenc}
\usepackage{apjfonts} 
\usepackage[figure,figure*]{hypcap}
\usepackage{longtable}
\usepackage{threeparttable}

\usepackage{amssymb}
\usepackage{amsmath}
\usepackage{fontawesome}
\usepackage{gensymb}
\usepackage{mathrsfs}
\usepackage{upgreek}
\usepackage{fontenc}
\usepackage{hyperref}
\usepackage{float}
\usepackage{tabularx}
\usepackage{booktabs}

\renewcommand{\d}{\ensuremath{\rm d}}

\graphicspath{{./}{figures/}}

\begin{document}
\title{Local Analogs of Primordial Galaxies: In Search of Intermediate Mass Black Holes with JWST NIRSpec 
}

\author[0000-0003-3152-4328]{Sara Doan}
\affiliation{George Mason University, Department of Physics and Astronomy, MS3F3, 4400 University Drive, Fairfax, VA 22030, USA}

\author[0000-0003-2277-2354]{Shobita Satyapal}
\affiliation{George Mason University, Department of Physics and Astronomy, MS3F3, 4400 University Drive, Fairfax, VA 22030, USA}

\author[0000-0003-3937-562X]{William Matzko}
\affiliation{George Mason University, Department of Physics and Astronomy, MS3F3, 4400 University Drive, Fairfax, VA 22030, USA}

\author{Nicholas P. Abel}
\affiliation{College of Applied Science, University of Cincinnati, Cincinnati, OH, 45206, USA}

\author[0000-0002-5666-7782]{Torsten B\"oker}
\affiliation{European Space Agency, c/o STSCI, 3700 San Martin Drive, Baltimore, MD 21218, USA}

\author[0000-0002-4375-254X]{Thomas Bohn}
\affiliation{ Hiroshima Astrophysical Science Center, Hiroshima University, 1-3-1 Kagamiyama, Higashi-Hiroshima, Hiroshima 739-8526, Japan}

\author[0000-0003-4693-6157]{Gabriela Canalizo}
\affiliation{Department of Physics and Astronomy, University of California, Riverside, 900 University Avenue, Riverside, CA 92521, USA}

\author[0000-0003-1051-6564]{Jenna M. Cann}
\affiliation{X-ray Astrophysics Laboratory, NASA Goddard Space Flight Center, Code 662, Greenbelt, MD 20771, USA}
\affiliation{Center for Space Sciences and Technology, University of Maryland, Baltimore County, 1000 Hilltop Circle, Baltimore, MD 21250}

\author[0000-0001-6697-7808]{Jacqueline Fischer}
\affiliation{George Mason University, Department of Physics and Astronomy, MS3F3, 4400 University Drive, Fairfax, VA 22030, USA}

\author[0000-0002-5907-3330]{Stephanie LaMassa}
\affiliation{Space Telescope Science Institute, 3700 San Martin Drive, Baltimore, MD 21218, USA}

\author[0000-0003-3229-2899]{Suzanne C.Madden}
\affiliation{Département d’Astrophysique, Université Paris-Saclay, Gif-sur-Yvette, France}

\author[0000-0002-0913-3729]{Jeffrey D. McKaig}
\affiliation{George Mason University, Department of Physics and Astronomy, MS3F3, 4400 University Drive, Fairfax, VA 22030, USA}

\author[0000-0001-7144-7182]{D. Schaerer}
\affiliation{Observatoire de Genève, Université de Genève, Chemin Pegasi 51, 1290 Versoix, Switzerland}
\affiliation{CNRS, IRAP, 14 Avenue E. Belin, 31400 Toulouse, France}

\author[0000-0002-4902-8077]{Nathan J. Secrest}
\affiliation{U.S. Naval Observatory, 3450 Massachusetts Avenue NW, Washington, DC 20392-5420, USA}

\author[0000-0003-0248-5470]{Anil Seth}
\affiliation{Department of Physics and Astronomy, University of Utah, 115 South 1400 East, Salt Lake City, UT 84112, USA}

\author{Laura Blecha}
\affiliation{University of Florida, Department of Physics, P.O. Box 118440, Gainesville, FL 32611-8440}

\author[0000-0001-8440-3613]{Mallory Molina}
\affiliation{Department of Physics \& Astronomy, Vanderbilt University, Nashville, TN 37235, USA}

\author[0000-0003-2283-2185]{Barry Rothberg}
\affiliation{U.S. Naval Observatory, 3450 Massachusetts Avenue NW, Washington, DC 20392, USA}
\affiliation{George Mason University, Department of Physics and Astronomy, MS3F3, 4400 University Drive, Fairfax, VA 22030, USA}

\correspondingauthor{Shobita Satyapal}
\email{ssatyapa@gmu.edu}

\begin{abstract}

\noindent  While JWST has delivered spectacular rest-frame optical/UV observations of galaxies at high-z, they lack the spatial resolution, sensitivity, and wavelength coverage necessary to identify the lowest mass black holes, constrain the ionizing radiation field, and map on parsec scales the interstellar medium (ISM). Local low metallicity galaxies with signatures of possible accretion activity are ideal laboratories in which to search for the lowest mass black holes and study their impact on the host galaxy. Here we present the first JWST NIRSpec IFS observations of  SDSS J120122.30+021108.3, a nearby ($z=0.00354$) extremely metal poor dwarf galaxy with no optical signatures of accretion activity but identified by WISE to have extremely red mid-infrared colors consistent with AGNs. These observations reveal a remarkable spectrum at exquisite spatial resolution, with over one hundred lines identified between $\sim$ 1.7-5.2 microns, an unresolved nuclear continuum source with an extremely steep spectral slope consistent with hot dust from an AGN ($F_\nu \approx\nu^{-1.5}$), and a plethora of \ion{H}{1}, \ion{He}{1}, and H$_2$ lines, with no lines from heavier elements, CO or ice absorption features, or PAHs. The H$_2$ lines include high-lying vibrational transitions, indicating the molecular gas is highly excited. Our observations reveal that the red WISE source arises exclusively from a bright central unresolved source ($<$ 3pc). Despite the signature of extremely hot dust in the unresolved nuclear source, there are no \ion{He}{2} lines or coronal lines identified in the spectrum, and, importantly, there is no evidence that the radiation field is harder in the nuclear source compared with surrounding regions. The near-infrared emission line spectrum of the central nuclear source is consistent with a young ($<$ 5 Myr) nuclear star cluster with stellar mass $\sim3\times 10^4$\,M$_\sun$. These observations reveal that either a metal poor stellar population can heat the dust to extremely high temperatures, a result that would have significant impact on the reliability of MIR color selection in AGN surveys and our understanding of the properties of the ISM and young stars in metal poor galaxies, or an accreting IMBH is deeply embedded and hidden even at near-infrared wavelengths.

\end{abstract}

\keywords{galaxies: active --- galaxies: Starburst --- galaxies: Evolution --- galaxies: dwarf --- optical: ISM --- line: formation --- accretion, accretion disks }

\section{Introduction}
Within its first year of operations, JWST has revolutionized our understanding of the high-z universe, discovering a population of low-mass, low-metallicity galaxies with vigorous star formation and extreme excitation conditions at unprecedented redshifts \citep[e.g.,][]{2023A&A...677A..88B, 2023A&A...677A.115C, 2023NatAs...7..622C, 2023NatAs...7..611R, 2023ApJ...955...54S,2024arXiv240518485C}. Among these discoveries, a remarkable population of low-metallicity star forming galaxies with broad lines consistent with faint Active Galactic Nuclei (AGNs) has been uncovered with redshifts up to 10.6, when the universe was only 430 Myr old \citep[e.g.,][]{2023arXiv230311946H,2023ApJ...954L...4K,2023ApJ...953L..29L, 2023arXiv230512492M, 2023arXiv230801230M, 2023A&A...677A.145U, 2024ApJ...964...39G, 2024ApJ...963..129M, 2024Natur.628...57F, 2024arXiv240302304W, 2024MNRAS.531..355U}. These faint AGNs may be part of a bigger population of compact galaxies with a unique "v-shaped" spectral energy distribution (SED) with a steep red continuum that have come to be known in the literature as "little red dots" (LRDs) \citep{2024ApJ...968...38K}. The newly discovered AGNs are significantly less luminous than previously uncovered in high redshift quasar surveys, and are inferred to be two orders of magnitude more numerous than the extrapolation of the UV-selected quasar luminosity function \citep{2024ApJ...963..129M}. The black holes appear to be relatively low mass ($\approx 10^6 - 10^8$\,$M_\odot$), much lower than found in previous high-z surveys \cite[e.g.,][]{2023ARA&A..61..373F}, yet overmassive relative to their host galaxies \citep[e.g.,][]{2023arXiv230311946H,2023arXiv230801230M, 2024arXiv240303872J,2024A&A...684A..24P}, although there are significant uncertainties in both black hole and stellar masses. The smallest mass black holes seem to be radiating at super Eddington rates \cite[e.g.,][]{2023arXiv230801230M,2023A&A...677A.145U,2023arXiv230812331P}, suggesting that they may be the missing link between black hole seeds formed at even higher redshifts and the first luminous quasars \cite[e.g.,][]{2023arXiv230805735F}.
\par
While the exquisite sensitivity of JWST has uncovered this remarkable population, broad line AGNs can only be identified for relatively massive black holes. For black hole masses less than $\approx 10^5$\,$M_\odot$, broadened lines associated with a central black hole can be too narrow and faint to be identified, leaving the long sought-after intermediate mass black holes (IMBHs) invisible to broad-line AGN searches. In addition,  Type 1 AGNs comprise only a small fraction of the total AGN population. Indeed, even amongst the population of AGNs in current surveys based on the strong optical narrow lines, Type 2 AGNs (i.e. the AGNs without visible broad lines) are about 4 times more numerous than Type 1 AGNs \cite{1995ApJ...454...95M}, and Type 1 and 2 AGNs in current optical surveys themselves capture only a small fraction of the total active galaxy population.
The dominant pathway for supermassive black hole (SMBH) formation is therefore still not robustly constrained \citep{2024ApJ...969...69L,2024arXiv240218773J, 2024ApJ...964..154P}.

As an added challenge, the most prominent lines detected at high-z are the rest-frame optical hydrogen recombination lines, and the strong optical narrow lines typically used as AGN diagnostics. Unfortunately, AGN diagnostics based on the strong optical narrow lines fail to identify and characterize the properties of low metallicity AGNs \cite[e.g.,][]{Groves2006, Cann2019}, resulting in optical narrow line ratios indistinguishable from star forming galaxies in JWST's high-z black holes \cite[e.g.,][]{2023arXiv230801230M}. In some,  but not all cases, semi-forbidden nebular lines tracing high density gas \cite[e.g.,][]{2023ApJ...953L..29L, 2023arXiv230512492M}  and faint high ionization forbidden lines \cite[e.g.,][]{2023ApJ...954L...4K, 2023arXiv230512492M, 2023arXiv230801230M, 2023A&A...677A.145U} are also detected. A few narrow line AGNs are now being discovered at high redshifts \citep{2023arXiv231118731S}, one of which may have a black hole mass between $\approx 10^5 -10^7$\,$M_\odot$ \citep{2024arXiv240218643C}. However, the sensitivity needed to identify IMBHs with masses less than $\approx 10^5$\,$M_\odot$ at high-z likely lies outside the reach of even JWST. Indeed, there is currently no robust evidence for a black hole with a mass between $\approx 150 - 10^4$\,$M_\odot$, a gap of roughly two orders of magnitude. This is a significant deficiency, since the mass distribution and host galaxy demographics of black holes in this mass range are needed to provide robust constraints on SMBH seed models, with some models predicting very early formation of SMBHs in dark matter minihalos and a genuine mass gap of black holes below $\approx 10^5$\,$M_\odot$ \cite[e.g.,][]{2024arXiv240709949C}. Moreover, the paucity of spectral lines, the lack of access to molecular features, and the poor spatial resolution of observations of high-z galaxies make it impossible to characterize in detail the AGN radiation field, and study its impact on the state and structure of the surrounding ISM.  As a result, crucial physics required to gain understanding of black hole-galaxy co-evolution, and the role accreting black holes play in cosmic reionization remain inaccessible. \\ 

While it is challenging to uncover IMBHs and study the detailed physics of their interaction with their host galaxies at high redshift, recent work has revealed a small population of metal-poor dwarf galaxies with extremely red mid-infrared colors consistent with hot dust heated by powerful AGNs and/or  high ionization lines, neither of which can by explained by any known stellar populations based on our current theoretical understanding \citep{2021MNRAS.508.2556I,2021ApJ...922..155M,Reefe2022, 2023arXiv230403726H, 2024ApJ...966..170H}. These local galaxies have stellar masses significantly below those accessible in the high redshift universe. Recent large-scale optical surveys also revealed that the emission lines with the highest ionization potential are preferentially found in lower mass galaxies, possibly indicating a hardening of the ionizing radiation field with decreasing stellar mass\citep{2022ApJ...936..140R, reefe2022class}. Since lower mass black holes produce a hotter accretion disk \citep{1973A&A....24..337S}, this finding can be consistent with accreting IMBHs powering the high ionization lines in the lowest mass dwarf galaxy population.  These findings hint at the possibility that objects like the progenitors of the faint AGNs discovered by JWST at high redshift {\it do} exist in the local universe, and can provide insight into the likely more common lower mass objects inaccessible even by JWST. However, while mid-infrared color selection has been shown to be highly successful at identifying large samples of AGNs \cite[e.g.,][]{2015ApJS..221...12S}, young stellar populations can potentially mimic AGNs in dwarf galaxies \citep{2016ApJ...832..119H}. Similarly, while high ionization lines are highly suggestive of accretion activity, most detections in the extremely metal poor population are of the \ion{He}{2}\,$\lambda$4686 (ionization potential (IP) = 54~eV) \citep{2012MNRAS.421.1043S} or [\ion{Ne}{5}]\,$\lambda$3426 (IP = 97~eV) \citep{2021MNRAS.508.2556I, 2023arXiv230403726H, 2024ApJ...966..170H} line. Alternative scenarios for the production of these lines include photoionization from extreme stellar populations, high mass X-ray binaries (HMXBs), or shock excitation from supernovae or stellar winds have also been proposed \citep{2012MNRAS.427.1229I, 2012MNRAS.421.1043S,2011A&A...536L...7I,2021MNRAS.508.2556I, 2021A&A...656A.127S}. Access to higher ionization potential lines (IP $>$ 100~eV), or so-called "coronal lines", are crucial to confirm AGNs in metal poor dwarf galaxies. Apart from AGN confirmation, \citet{Cann2018} has shown that infrared coronal lines are expected to be enhanced in the spectra of accreting IMBHs compared with higher mass BHs, and can potentially be a powerful probe of the black hole mass in AGNs. and potentially constrain the black hole mass \citep{Cann2018}.

In this work, we present JWST NIRSpec Integral Field Spectroscopy (IFS) observations of SDSS J120122.30+021108.3 (hereafter J1201+0211), a nearby (z=0.00354) low metallicity dwarf galaxy three orders of magnitude less massive than the LMC. It is one a handful of low metallicity dwarfs known in the local universe with extreme mid-infrared colors consistent with AGNs \cite[e.g.,][]{2011A&A...536L...7I}, and optical narrow line ratios consistent with JWST's newly discovered high-z AGNs, making it an ideal laboratory in which to search for an IMBH in a completely uncharted mass regime, and to study in unprecedented detail its impact on the host in an environment similar to the high-z AGNs newly discovered by JWST. Because of their proximity and the resulting potential for high sensitivity spectroscopy that is impossible to obtain at the very highest redshifts, these galaxies are also ideal laboratories for providing insight into the first generation of stars, black holes, and the ionizing photons that contributed to cosmic reionization. If current black hole-galaxy scaling relations hold in this mass range, the central black hole will be significantly less massive than any currently known. The primary goal of this work is to search for near-infrared high ionization lines in order to confirm the presence of an accreting IMBH. We note that there are high ionization lines in the optical and ground-accessible near-infrared wavelength range, but based on a review of the literature, these shorter wavelength lines are as much as two orders of magnitudes fainter than the high ionization lines longward of $\approx2.4\mu\mathrm{m}$, where the sky brightness and atmospheric absorption precludes detection from the ground. Moreover, coronal lines accessible by NIRSpec extend up to $\approx500$~eV, far greater than the ionization potential that has been detected to date in any dwarf galaxy. In this work, we also present XMM-Newton observations of J1201+0211. To estimate the distance to J1201+0211, we assume a flat $\Lambda$CDM cosmology with $H_0=70$~km~s$^{-1}$~Mpc$^{-1}$, and $\Omega_\mathrm{M} = 0.3$, and $\Omega_\Lambda = 0.7$. The resulting luminosity distance is $D_L=15.2$\,Mpc, consistent with that obtained using the Cosmicflows-3 local velocity flow model \citep{2019MNRAS.488.5438G,2020AJ....159...67K}

\section{Target selection}
J1201+0211, which lies at the southern edge of the Virgo Cluster, is an extremely metal poor (XMP) dwarf galaxy (defined as galaxies with metallicities less than 10\% Solar), first identified in one of the earliest catalogs of XMPs based on the Sloan Digital Sky Survey (SDSS) \citep{2003ApJ...593L..73K}. XMP galaxies, an extremely rare population, are typically found within blue compact dwarf galaxies (BCDs) — a class of galaxies characterized by compact morphology, strong emission lines, and a blue continuum indicating intense star formation \citep{1981ApJ...247..823T}. The gas phase metallicity of J1201+0211 is $12+\log(\mathrm{O}/\mathrm{H})$ = $7.45 \pm 0.03$ \citep[or $\simeq$6\% the Solar value assuming a Solar metallicity of 12 + log(O/H) of 8.69;][]{2006NuPhA.777....1A}, comparable to those of the faint broad line AGNs discovered by JWST, the star formation rate based on the Balmer lines is $1.97\times10^{-2}$\,M$_\sun$\,yr$^{-1}$,  and its stellar mass based on fits to the SDSS broadband photometry is $\log(M_*/M_\odot) = {6.09}^{+0.44}_{-0.25}$ \citep{2016ApJ...827..126B}. We checked the deep DESI Legacy Imaging Survey and found no evidence of extended emission from older stars that could dominate the mass as is seen in some BCDs \citep{2013MNRAS.431..102M}. There is a bright central continuum source seen in the optical SDSS image with colors characteristic of a young stellar population, with a faint chain of compact sources discernible after an image stacking analysis out to $\sim$ 1.4~kpc from the central source \citep{2008A&A...491..113P}. \citet{2007A&A...464..859P} suggest that its somewhat disturbed morphology is possibly suggestive of an interaction with a very faint companion, although no clear companion or evidence of tidal tails is seen in the DESI Legacy Survey images. Archival UV spectroscopy from HST/COS reveals prominent nebular emission in the resonant \ion{C}{4}\,$\lambda$1548,1550 doublet, exceedingly rare in nearby low redshift star forming galaxies but found in high redshift Ly$\alpha$ emitters \citep{2015MNRAS.454.1393S, 2017ApJ...839...17S}; as well as the $\lambda$1640 \ion{He}{2} line, indicating the presence of highly ionized gas. Like the majority of metal poor BCDs, J1201+0211 has a high gas fraction, and little evidence for rotation based on HI profiles \citep{2016MNRAS.463.4268T}.

\input{panels/BPTWISE}

There is no evidence for an AGN based on its optical narrow emission line ratios in the SDSS aperture (3\arcsec), using the Baldwin-Phillips-Terlevich (BPT) diagnostics \citep{1981PASP...93....5B}. As can be seen in the left panel of Figure~\ref{fig:bptwise}, its location in the upper left corner of the BPT diagram is consistent with low metallicity star forming galaxies, as well as the high-z AGNs recently discovered by JWST \cite[e.g.,][] {2023arXiv230801230M}. There is no clear detection of the \ion{He}{2}$\lambda$4686 line or any optical coronal line in the SDSS spectrum. 

Based on the AllWISE catalog\footnote{\url{https://wise2.ipac.caltech.edu/docs/release/allwise/}}, the [3.4$\mu$m] - [4.6$\mu$m] (hereafter $W1-W2$) color of J1201+0211 is  $1.835  \pm 0.040 $, extremely red for a starforming galaxy. Mid-infrared color selection has been shown to be a very reliable technique in identifying obscured and unobscured AGNs in large scale surveys \cite[e.g.,][]{2015ApJS..221...12S, 2019ApJ...876...50L}. This is because AGNs can heat the dust to high temperatures, produces a strong mid-infrared continuum and an infrared SED that is clearly distinguishable from normal star-forming galaxies for both obscured and unobscured AGNs in galaxies, where the emission from the AGN dominates over the host galaxy emission  \cite[e.g.,][]{2007ApJ...660..167D, 2012ApJ...753...30S, 2013MNRAS.434..941M}. In particular, at low redshift, the $W1-W2$ color of galaxies dominated by AGNs is considerably redder than that of inactive galaxies \citep[see Figure 1 in][]{2012ApJ...753...30S}. The $W1-W2$ color of J1201+0211 is  well above the widely used color cut ($>$0.8) used to identify AGNs by \citet{2012ApJ...753...30S}. While there are less than a handful of dwarfs with similar $W1-W2$ colors, the $W2-W3$ colors are typically much redder \citep{2011ApJ...736L..22G}. The $W2-W3$ color, on the other hand for J1201+0211 is $3.984  \pm 0.040 $, which is much more consistent with the colors of AGNs. Based on a theoretical investigation of the mid-infrared spectral energy distribution (SED) produced by dust heated by an AGN and an extreme metal poor starburst, \citep{satyapal2018} put forward a 3-band color cut that excludes even the most extreme stellar poor stellar populations based on our current understanding (red line in Figure~\ref{fig:bptwise}). As can be seen, the WISE colors of J1201+0211 falls within the \citet{satyapal2018} bounding box, indicating that its WISE colors cannot easily be explained through dust heating from extreme stars, based on our current understanding.   We note that MIR color selection only selects the most dominant AGNs \citep{satyapal2018}, which comprise only a small fraction of the total population, so J1201+0211 is an extreme case of a possibly more numerous population of AGNs in the metal poor dwarf galaxy population.  Based on an inspection of the AllWISE/NEOWISE multi-epoch catalogs, J1201+0211 does not show any signs of MIR variability. We note however that variability selection of AGNs in dwarf galaxies is ineffective unless higher-cadence data are used \citep{2020ApJ...900...56S}, with only a small fraction of optically identified AGNs displaying MIR variability based on AllWISE/NEOWISE multi-epoch photometry\citep{2024arXiv240800666A}.

In the right panel in Figure~\ref{fig:bptwise}, we show the WISE color-color diagram for J1201+0211 compared to that of the general dwarf population. As can be seen, J1201+0211 is extremely rare. Indeed, only 0.004\% of the entire sample of star forming galaxies in the SDSS DR8 catalog have $W1-W2$ > 1.8. Even amongst the 151 optically identified AGNs in the \citet{2013ApJ...775..116R} sample, none have  $W1-W2$ > 1.8.  In Figure~\ref{fig:bptwise}, we also show the location of the  [\ion{Ne}{5}] emitting dwarfs from \citep{2021MNRAS.508.2556I}, which includes the star forming XMP dwarf SBS 0335-032, recently found to show MIR variability consistent with an AGN \citep{2023arXiv230403726H}. As can be seen, most do not show as extreme mid-infrared colors to J1201+0211, despite the presence of a hard ionizing radiation. Even amongst the sample of 82 starforming galaxies with strong $\lambda$4686 \ion{He}{2} emission identified by \citet{2012MNRAS.421.1043S}, only one has $W1-W2$ > 1.8. The extreme nature of this source, coupled with its extremely low metallicity and low stellar mass, makes J1201+0211 an ideal target in which to search for an IMBH with JWST. Based on its stellar mass, this target can potentially unveil an IMBH of $\approx1000$\,$M_\odot$, roughly an order of magnitude lower than any other nuclear black hole known in a galaxy thus far.

\section{JWST NIRSpec Observations and Data Reduction}
J1201+0211 was observed with  NIRSpec IFS \citep{2022A&A...661A..82B, 2022A&A...661A..80J} on 6 June 2023 as part of GO program 1983 (PI: Satyapal). Data was collected using the NIRSpec IFU medium resolution observations in the G235M/F170LP and G395M/F290LP grating/filter combinations using the NSIRS2RAPID readout with 60 groups and 2 integrations at four dither positions resulting in a total on-source exposure time of $\sim\textrm{7100}$~s for each grating/filter setting. The resulting wavelength coverage obtained with our choice of grating/filter combinations was $\sim 1.6 - 5.2~\mu$m with a nominal resolving power of R $\sim\textrm{1000}$. To maximize observing efficiency, we obtained a single LeakCal exposure at only one position with the same integration time as our observation in order to remove the effects of leakage from any permanently open microshutters. The $3\arcsec\times3\arcsec$ field of view (FOV) of the NIRSpec IFS corresponds to $\approx 200\times200$~pc at the distance to J1201+0211.
\par
We downloaded the raw data files from the Barbara A. Mikulski Archive for Space Telescopes (MAST) and subsequently processed them with the JWST Science Calibration pipeline version 1.12.5 \citep{2022zndo...7229890B} under the Calibration Reference Data System (CRDS) context $\textrm{jwst}\_1100.pmap$.  The level 1 files downloaded from MAST were processed using the Detector1 stage (stage 1), which performs detector-level corrections and generates count-rate images. We applied several modifications to the default data reduction steps to improve data quality and to remove spurious features or recover real features that were removed using the default outlier rejection algorithm. In order to remove artifacts generated by cosmic-ray hits on the detector, we applied the snowball flagging for the jump step during the first stage of the pipeline. We also removed the 1/f noise, which appeared as correlated vertical bands in the count rate images, by implementing the NSClean algorithm \citep{2024PASP..136a5001R}, which performs and subtracts a polynomial fit (after removing bright pixels from the source through a sigma clipping algorithm).  The stage2 images were then resampled and combined into a final data cube through the Calwebb\_spec3 processing (stage3) using the ‘drizzle’ method with a 0.1$\arcsec$ spaxel size. To enhance the spatial resolution of the data cubes, we also generated cubes with a spaxel size of 0.05$\arcsec$ (corresponding to $\sim3$~pc per spaxel). The resulting data cubes, particularly for the smaller spaxel size, display the well-known sinusoidal modulations in the spectrum caused by the undersampling of the point-spread function (PSF). To correct these so-called "wiggles", we applied our own custom-code similar to that developed by \citet{2023A&A...679A..89P} that masks out the detected lines and models and subtracts the sinusoidal variations (see Matzko et al. 2024, in prep). Note that the "wiggles" are insignificant in spectra extracted from apertures larger than $\sim$0.2$\arcsec$ radius. 

\subsection{Line Identification and Spectral Fitting Procedure}

Given the multitude of lines detected in the spectrum, and the possibility of spurious features remaining in the processed cubes, line identification requires a careful methodology. Since the main science goal of the program was to search for evidence for an IMBH through the presence of highly ionized gas, this is particularly critical to ensure that any reported coronal lines are robustly identified. To ensure features are genuinely identified spectral lines, and to provide additional validation of the final data products, the extracted spectrum centered on the nucleus with a 0.4 $\arcsec$ radius aperture size was carefully examined in each of the four individual dither positions. The nucleus was defined as the brightest central region seen in both the continuum and ionized gas, as traced by the hydrogen recombination lines. Any feature that did not appear in all four dithers was flagged and removed. Similarly, features that appeared in all dither positions but were removed by the default outlier rejection algorithm were retained. We note that the default pipeline products downloaded from MAST contained a significant number of spurious features that needed to be removed after careful inspection, some of which were centered on coronal line wavelengths. A significant number of genuine features were also found to have been removed by the default outlier algorithm. 

\par

Spectral fitting was performed on all identified spectral features and is done, as in our previous work, with the open-source Python code Bayesian AGN Decomposition Analysis for SDSS Spectra (\textsc{badass})\footnote{\href{https://github.com/remingtonsexton/BADASS3}{https://github.com/remingtonsexton/BADASS3}} adapted to apply to the JWST spectral region. A full description of the code can be found in \citet[][]{sexton_2020}. Initial fitting with BADASS is performed using a likelihood maximization routine from SciPy, followed by a Markov Chain Monte Carlo (MCMC) routine using the affine-invariant Emcee sampler \citep{2013PASP..125..306F} which yields robust uncertainties. Gaussian profiles were employed for features with a third order Legendre polynomial fit for the continuum. Spectral fitting across the full spectrum was applied on each individual spaxel from which flux, FWHM, and velocity offset maps were generated for all identified features. Because of the plethora of lines in the high sensitivity spectrum, and the possibility that spectral features can be shifted due to outflows, as is often seen particularly for coronal lines \citep[e.g.,][]{2006ApJ...653.1098R,2011ApJ...743..100R, 2018ApJ...858...48M}, we adopted a careful procedure in identifying lines. This included first selecting the closest atomic and molecular lines to each feature, followed by a comparison with predictions based on photoionization models with low metallicity stellar populations described in section 5.6.1. In addition, we visually inspected flux maps and noted that the morphology was distinctly different for atomic and molecular lines. In several instances, the closest line identified was a molecular hydrogen line; however based on model predictions and the morphology of the flux map, the line was clearly a hydrogen recombination line with a wavelength in very close proximity to that of the molecular hydrogen line. The combination of visual inspection of the line flux maps and comparison to model predictions for the various lines was used to generate the final list of spectral features identified in the nuclear spectrum. In addition to generating spectral maps, the spectra from a 0.4$\arcsec$ radius aperture centered on four separate locations sampling the FOV were extracted and spectral fits performed on all features in the extracted spectra.

\input{tables/cl_upper_limits}

\section{XMM-Newton Observations}

J1201+0211 was observed by XMM-Newton on June 12-13, 2021 (Obs. ID 0881560901, PI: Satyapal), with an exposure time of 34 ks. Data were accessed through the XMM-Newton Archival Database, and reprocessed using version 19.1.0 of the XMM Science Analysis Software (SAS) \citep{gabriel2004} using the corresponding Current Calibration Files (CCFs). The data was reprocessed using a custom automated pipeline\footnote{https://github.com/thatastroguy/XMM-Processing} \citep{Pfeifle2023} following the steps in the XMM ABC Guide\footnote{https://heasarc.gsfc.nasa.gov/docs/xmm/sl/intro.html}. In short, the pipeline builds a CCF index file (\textsc{cifbuild}), creates the Observational Data File (ODF) summary file (\textsc{odfingest}), runs the \textsc{emproc} and \textsc{epproc} tasks to reprocess the data, and applies standard filters (PATTERN $\leq 4$ and FLAG == 0). The observation was affected by significant background flaring, so we examined the light curve extracted with PATTERN == 0 and photon energies $\geq$ 10000. The light curves were dominated by background flares in the last $\approx5-10$ks of the exposure, resulting This in a mean count rate of 1.58 and 0.21  counts s$^{-1}$ for the pn and MOS cameras, respectively. We filtered based on an upper count rate limit of 0.4 and 0.2 counts s$^{-1}$ for the pn and MOS cameras, respectively, resulting in a good time interval (GTI) of 24 ks. A full-band (0.3 -10 keV) image, binned at the native pixel size of 4.1$\arcsec$ was constructed using \textsc{evselect}. The data showed a non-detection for J1201+0211, so an upper limit was derived using the \textsc{eregionanalyse} routine to extract count rates and 3$\sigma$ upper limits for a circular source region with a radius of $30\arcsec$.

\section{Results and Discussion}
\subsection{Morphology of the Continuum, Ionized and Molecular Gas}

\begin{figure*}
\centering
\includegraphics[width=0.95\textwidth]{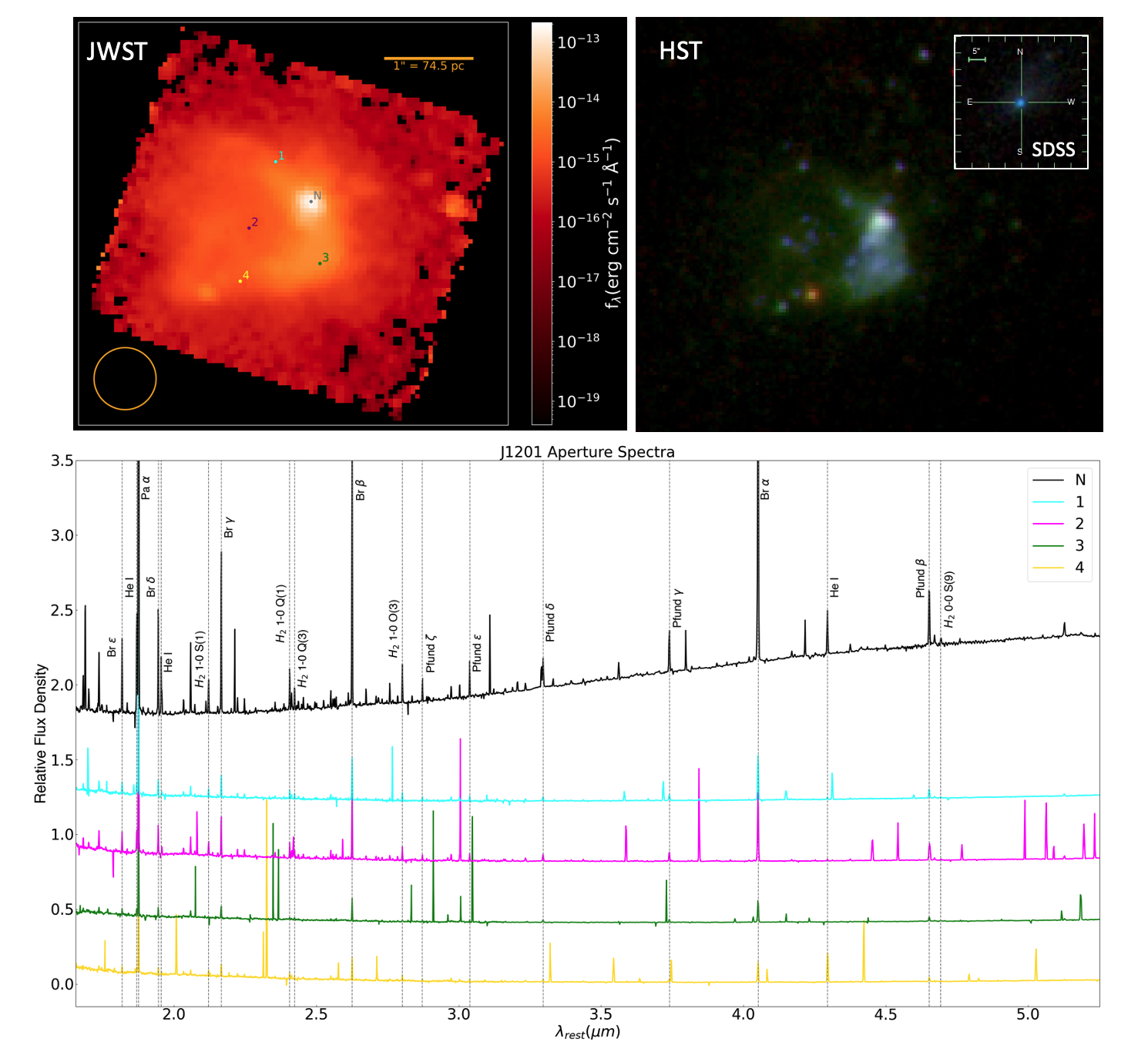}\\
\caption{(Top:) JWST NIRSpec IFU Pa$\alpha$ image(left) and HST WFC3/UVIS broad band 3-color image (right) of J1201+0211. The SDSS image is shown in the upper right. The JWST image shown is for the 0.05\arcsec spaxel scale, which corresponds to 3.65 pc at the distance of J1201+0211.(Bottom:) Color-coded extracted 1-D IFU spectra are displayed from 5 apertures shown in the JWST image. The yellow circle in the top left image denotes the size (radius = 0.4\arcsec) of the extraction aperture for the spectra. Vertical dotted lines correspond to the locations of a few prominent lines.}
\label{fig1}
 \label{fig:spectra_regions}
\end{figure*}

 The JWST observations reveal a spectrum with unprecedented sensitivity at exquisite spatial resolution, with over one hundred lines detected between $\approx$ 1.7-5.2~$\mu$m.  In Figure \ref{fig:spectra_regions}, we display the  Pa$\alpha$ image alongside the HST broadband image of the central few arcseconds of the galaxy. The SDSS image is also displayed in the upper right. As can be seen, the enhanced spatial resolution observations reveal intricate structures on parsec scales that are not seen in the SDSS image. The galaxy shows a bright central source, which we refer to as the nucleus, several additional fainter clumps with diffuse emission and a shell-like structure connecting the clumps. A central cavity is seen in the ionized gas to the east of the nucleus.  In Figure~\ref{fig:flux_maps}, the line-free continuum emission at 2.2~$\micron$ and 4.6~$\micron$ is displayed in the upper panels. The central source dominating the emission in the Pa$\alpha$ image coincides with the peak in both continuum maps. As can be seen, the bright nuclear source also dominates the emission in the 2.2~$\micron$ continuum map, with a fainter source seen in the south east. This south-east source disappears at longer wavelengths, as can be seen in the upper right panel in Figure~\ref{fig:flux_maps}. Based on the radial profile of the continuum integrated over the WISE W2 band (centered at 4.6 ~$\mu$m) in the 0.05\arcsec spaxel scale data, the nuclear source is unresolved, indicating that its extent is less than $\approx$ 3~pc.  These observations demonstrate that the red colors identified by WISE are associated with a bright unresolved nuclear source.
 \par
 In the bottom left panel of Figure~\ref{fig:flux_maps}, we show the extinction map obtained using the Pa$\alpha$ and Br$\alpha$ flux ratio, assuming Case B recombination (see Section 5.3 for details). The extinction map reveals a patchy distribution, with significant extinction in the north east.  The peak extinction is seen North of the nuclear source. The ionized cavity observed in the Pa$\alpha$ image therefore does not appear to be caused by extinction, but is a genuine cavity in the ionized gas. The multiphase ISM in Dwarf Irregulars is often found to be characterized by holes and shell-like structures, thought to be created by winds from massive stars and supernovae \cite[e.g.,][]{2019A&A...626A..23C}. The hard interstellar radiation field produced by massive stars, combined with the low dust content and shallow potential well typically associated with dwarf galaxies, is thought to carve out cavities in the ISM. The increase in porosity of the ISM at low metallicity is of particular relevance in reionization era galaxies, since it could facilitate the escape of ionizing photons \citep{2016ARA&A..54..761S}. 
 \par

In the bottom right panel of Figure~\ref{fig:flux_maps}, we show the flux map of the brightest molecular hydrogen line in the spectrum. As can be seen, the $H_{2}$ 1-0 Q(1) 2.4$\micron$ map shows a distinctly different morphology from the ionized gas maps. In addition to the nuclear source, clumps of emission are seen in the south, where no corresponding clumps of emission in the ionized gas are seen. The $H_{2}$ line ratios throughout the field of view vary considerably indicating that the excitation of the molecular gas changes significantly within the central $\sim$ 200~pc.  At low metallicity, because of the harder interstellar radiation fields and lower shielding by dust, CO, the usual proxy for cold molecular hydrogen, tends to be photodissociated, resulting in CO dark $H_{2}$ gas \citep{1988ApJ...334..771V,2013ARA&A..51..207B}. The ro-vibrational transitions of $H_{2}$ in the near-infrared probe warm molecular gas. The warm molecular gas traced by the $H_{2}$ lines may be one of the only avenues of investigating the state and structure of the molecular gas in very low metallicity environments. Recent JWST observations of M33 have shown that the warm molecular gas may constitute over 75\% of the total molecular gas mass \citep{2023ApJ...948..124H}, a much higher fraction than previously thought \cite[e.g.,][]{2007ApJ...669..959R}. Detailed photodissociation region (PDR) models will be carried out in a future paper. The plethora of lines detected by JWST in J1201+0211 will provide a benchmark for PDR models in a metallicity regime well below existing studies in the LMC and SMC.

\input{panels/flux_maps}

\subsection{Identified Spectral Features}

In Figure \ref{fig:spectra_regions}, we show the full spectrum extracted from a 0.4$\arcsec$ radius aperture centered on the nucleus and four additional regions that sample the FOV. The aperture size and locations are displayed in the upper left panel of Figure \ref{fig:spectra_regions}. As can be seen, there is a steeply rising slope only in the unresolved nuclear source, with all other apertures displaying a flat slope typically seen in young star forming regions. The continuum emission beyond $\sim$ 3~$\micron$ is well-fit by a power law with a power law index  consistent with hot dust in AGNs ($F_\nu \approx\nu^{-1.5}$). As can be seen, the spectrum is dominated by hydrogen recombination lines, ro-vibrational lines of molecular hydrogen, and \ion{He}{1} lines. The multitude of H$_2$ lines include high-lying vibrational transitions, indicating that the molecular gas is highly excited. We also note that the identified spectral features and line ratios vary throughout the FOV, indicating variable ISM conditions within the central $\sim$ 200~pc. The striking number of lines detected, and the variation in identified features in each spaxel highlights the challenges associated with identifying featureless regions of the spectrum to fit the continuum, and performing an automated data analysis methodology.
\par
 
In Table 2 in the Appendix, we list all identified features in the nuclear aperture along with line profile fit parameters. Because the observations presented here represent the first near-infrared spectrum obtained by JWST in primordial galaxy analog in this metallicity regime, we provide a comprehensive list of all visually identified features. The final list was obtained by careful inspection in all four dither positions performed independently by multiple team members, some of which are not 3$\sigma$ detections. Note that in this paper, we list spectral features identified only in the nuclear aperture. We note that additional lines, particularly from molecular hydrogen, are seen in the other apertures but not in the nuclear aperture. Spectral fits covering wavelength windows of the spectrum with significant lines in the nuclear aperture are displayed in Figure \ref{fig:spectra_sample}. The brightest line in the nuclear aperture is the Pa$\alpha$ line. There are a total of 27 hydrogen recombination lines, 60 molecular hydrogen lines, and 9 \ion{He}{1} lines. No broad lines are observed in the spectrum, and there is no evidence for circular rotation of the gas. The velocity shifts are generally $\lessapprox$ 50~km/s across the FOV for most detected features. There are no \ion{He}{2} lines detected in the spectrum, and no coronal lines. The upper limits for the brightest \ion{He}{2} line in this spectral region, along with common coronal lines detected in AGNs are listed in Table \ref{tab:cl_ident}. In order to determine the upper limits, we construct a window around the emission line of interest, centered at the emission line wavelength and extending 0.003 microns on either side. The continuum on each side of the line window is sampled until at least 5 good channels on each side are included in the analysis; all emission lines are masked from the data to ensure we are only sampling the surrounding continuum. We then integrate over a Gaussian profile whose amplitude is equal to the standard deviation in the continuum windows and FWHM is equal to the NIRSpec instrumental resolution in the emission line window to find the upper limit of the emission line. Indeed, there are no features identified from atoms other than hydrogen and helium in this metal poor dwarf in any of the apertures. We also do not detect any PAH features. It is well known that PAH emission is depressed in metal poor galaxies  \cite[e.g.,][]{2005ApJ...628L..29E, 2006ApJ...641..795O, 2006A&A...446..877M, 2010ApJ...712..164H}. We searched the entire FOV and find no evidence for PAH emission across the entire central $\sim$ 200$\times$200~ pc  region. There are also no signatures of the CO band head typically seen in the spectral of cool stars, or ice absorption features.

\input{panels/sample_spectra}
  
 \subsection{Extinction Toward the Ionized Gas}
 
We estimate the extinction toward the ionized gas using the Pa$\alpha$ and Br$\alpha$ flux ratio assuming Case B recombination with an electron temperature $T _{e}$ = 10,000~K and electron density $n _{e}$=100~cm$^{-3}$. We use the calculated intensity ratios of the hydrogen recombination lines from \citet{1987MNRAS.224..801H} for the intrinsic line ratios. Note that because the Pa$\alpha$ (1.875~$\micron$) and Br$\alpha$ (4.051~$\micron$) lines are widely separated in wavelength, the relative line intensities will be affected by the greatest differential reddening amongst the suite of detected lines. They are also the brightest lines detected in the spectrum. The choice of these lines therefore enables the highest sensitivity flux map from which the extinction can be analyzed throughout the central tens of parsecs. In our analysis presented here, we assume a foreground screen model for the extinction and an extinction power law of $\tau _{\lambda }\propto \lambda ^{-1.85}$ often adopted for the near-infrared \citep{2009ApJ...699.1209F}. We then use the ratio of infrared to optical extinction derived from recent JWST observations of 30 Doradus from \citet{2023A&A...671L..14F} (see their Table 4). The resulting extinction map shown in the bottom left panel of Figure~\ref{fig:flux_maps} reveals significant extinction, with a peak value $A_{v}$ $\sim$ 8 mag north of the nuclear source. The relatively high level of extinction is unusual for dwarf galaxies, which generally display low levels of extinction \cite[e.g.,][]{2008A&A...486..393V, 2009ApJ...703.1984I, 2016MNRAS.457...64I}. 

We note that the extinction curve and dust geometry along the line of sight can vary throughout the FOV. Numerous studies have found steeper slopes in regions with massive young stars, possibly as a result of destruction of smaller grains from the ionizing radiation field. Using the suite of hydrogen recombination lines detected in J1201+0211, a near-infrared extinction curve can be derived for the first time at this metallicity regime on parsec scales, providing insight into the dust properties and dust distribution in primordial galaxy analogs. We note that preliminary inspection of the extinction curve reveals possible variations in the power law slope as a function of position in the galaxy, possibly indicating changes in the grain size distribution or geometry of the dust distribution along the line of sight. Our data also enables an analysis of the extinction toward the \ion{He}{1}  as well as the $H_{2}$ emitting gas. Using the suite of lines, the extinction curve can be derived toward the ionized and molecular regions. The comparative analysis may give important insight into the different stages of dust clearing, and enable an understanding of the properties and spatial distribution of deeply embedded young stars, an analysis that cannot be done directly at high redshift. We note that the PSF is under-sampled in NIRSpec and varies across the wavelength range for the two recombination lines; the extinction map therefore can contain artifacts resulting in the varying PSF. A thorough analysis of the extinction and extinction law that takes into account PSF effects will be presented in a future paper. We also note that the extinction obtained using the NIRSpec data only samples the extinction toward the gas emitting the near-infrared emission lines; the possibility of a deeply embedded source invisible even in the near-infrared cannot be determined without longer wavelength observations. 

\subsection{Lack of High Ionization lines}

No high ionization lines were detected in the near-infrared spectrum of J1201+0211 despite the exquisite sensitivity of JWST. In Figure~\ref{fig:CL_ULs}, we plot the flux upper limits of the [\ion{Si}{6}] 1.962~$\micron$ line, the mostly commonly detected near-infrared coronal line in local AGNs, and the [\ion{Ca}{8}] 2.32$\micron$ line verses the WISE W2 (4.6~$\micron$) flux 
\input{panels/coronal_line_ULs} for J12011+0211, in comparison with the values for the hard X-ray selected Swift/BAT sample relation from \citet{2017MNRAS.467..540L}. The solid line represents the best fit linear regression for the sample of hard X-ray selected AGNs, and the dashed lines represent the 1$\sigma$ scatter in the sample.  In each plot, the value of the extinction at the wavelength of each line required to move the upper limit up to the lower bound in the local relations (yellow triangle) is listed in the legend. As can be seen the flux upper limits obtained from JWST are highly sensitive and imply that if the putative AGN in J1201+0211 is similar to well-known more massive AGNs, the coronal lines should have been detected, unless it is deeply embedded, invisible even at near- infrared wavelengths.

\subsection{ X-ray Results}

Using the source and background extraction regions described in Section 4, we find an upper limit to the number of counts in the 0.3 -10 keV band of 0.002 counts~s$^{-1}$. Assuming a power law X-ray spectrum with a power law index of $\Gamma$ = 1.7, the 3$\sigma$ upper limit to the 2-10~keV flux assuming no intrinsic absorption is 2.85 $\times10^{-14}$\,erg\,cm$^{-2}$\,s$^{-1}$. Assuming a distance of D=15.2~Mpc, the corresponding upper limit to the luminosity in the hard band is $L_{2-10\,\mathrm{keV}}\sim 5\times10^{38}$\,erg\,s$^{-1}$.

\subsection{The Nature of the Nuclear Source}
\subsubsection{A Nuclear Star Cluster?}
We explored whether the unresolved nuclear source could be consistent with a nuclear star cluster (NSC). Given the spatial resolution limits ($<$ 3~pc), this possibility cannot be ruled out. Indeed, based on the measured half-light radii of young star clusters in over 6,000 clusters in 31 nearby galaxies, \citet{2021MNRAS.508.5935B} find that galaxies share a common cluster radius distribution, with the peak at around 3~pc. NSC sizes range from $\sim$ 1-20~pc, with a median size of 4.5~pc in late-dwarfs in galaxies with stellar masses below $\sim 10^7$\,M$_\sun$ \citep{2020A&ARv..28....4N}. While NSCs are ubiquitous in higher mass galaxies, with an occupation fraction as high as $\sim$ 80\% in early-type dwarfs with stellar masses $\sim 10^9$\,M$_\sun$, the occupation fraction drops to near zero at stellar masses $\sim 10^6$\,M$_\sun$ \citep[see review by][]{2020A&ARv..28....4N}, and even lower still in late-type dwarfs \citep{2021MNRAS.507.3246H}. With the suite of hydrogen recombination lines, it is possible to calculate the total ionizing photon flux produced by the nuclear source, and to estimate the stellar population needed to replicate it. Using the extinction derived from the Pa$\alpha$ and Br$\alpha$ flux ratio, the extinction-corrected Pa$\alpha$ flux from the nuclear aperture is $3.88 \pm 0.22\times10^{-15}$\,erg\,cm$^{-2}$\,s$^{-1}$. Assuming that none of the photons escape the nebula, and there is no absorption of ionizing photons by dust, the total ionizing photon flux is given by

\begin{equation}
Q(H)=\frac{4\pi D^{2}F_{Pa\alpha }\alpha _{B}}{h\nu _{Pa\alpha }\alpha _{Pa\alpha }^{eff}}
\end{equation}

where D is the distance to J1201+0211, $h\nu _{Pa\alpha }$ is the energy of a Pa$\alpha$ photon, and $\alpha _{B}$ and $\alpha _{Pa\alpha }^{eff}$ are the total hydrogen and effective  Pa$\alpha$ Case B recombination coefficients. We use the calculated intensity ratios of the hydrogen recombination lines from \citet{1987MNRAS.224..801H}, assuming an electron temperature $T _{e}$ = 10,000~K and electron density $n _{e}$=100~cm$^{-3}$. Using a distance of D=15.2~Mpc, we obtain an ionizing photon flux of Q(H) $\sim  3\times10^{51}$\,s$^{-1}$, comparable to the total ionizing flux (r$<$ 10~pc) from the giant star cluster R136 in 30 Doradus in the LMC \citep{1998MNRAS.296..622C}. 
\par

Because there are no high ionization lines present and the spectrum is dominated almost exclusively by hydrogen lines, it is difficult to constrain the radiation field and physical conditions of the ionized gas. However, because the ionization potential of $He^{+}$ (24.6~eV) is higher than that of $H^{+}$, the detection of the \ion{He}{1} lines can be used to gain some insight into the hardness of the radiation field. We compared the JWST spectrum of J1201+0211 to that predicted from gas ionized by a stellar population using the spectral synthesis code \textsc{Cloudy} version c17 \citep{ferland2017}. For the stellar ionizing radiation field, we used the instantaneous burst single stellar population model  from the BPASS 2.1 \citep{2017PASA...34...58E}, which follows the evolution of single and binary stars from the STARS stellar evolution tracks \citep{2008MNRAS.384.1109E}. Because there are no high ionization lines and no \ion{He}{2} emission detected, we adopt here the single burst model. The inclusion of binary evolution hardens the radiation field and increases the stellar contribution to the $He^{++}$ ionizing flux, an effect that is unnecessary to adopt given the observed spectrum. We assume a Salpeter initial mass function (IMF) of power-law index $\alpha=-2.35$ \citep{1955ApJ...121..161S} and an instantaneous burst, with lower and upper mass cutoffs of 1 and 100 $M_\sun$, respectively. We adopt a one-dimensional spherical model with a closed geometry, and allow the age to vary from 1 to 20 Myr.  We set the inner radius of the cloud and the total luminosity such that the ionization parameter at the inner face of the cloud is $U$ $=10^{-2}$ for the 10~Myr population, which is a typical value based on observations of the optical BPT emission lines in local galaxies and H II regions \citep{dopita2000, moustakas2010}. Note that since the input stellar radiation field softens with age, reducing the number of ionizing photons, the ionization parameter for our model varies with age.  The ionization parameters in our model, for the chosen SED, varies from $U$ $=10^{-0.3}$ to  $U$ $=10^{-3.4}$.  The higher ionization parameter models can lead to a significant number of ionizing photons being absorbed by dust, although since we are dealing with a low metallicity environment, this effect will be less important.  Note that we adopt a stellar metallicity one tenth solar and we scale all metals along with the dust abundance by a factor of 0.10.  All calculations are stopped at the ionization front, defined here to be when the fraction of ionized hydrogen, $H^{+}$$/$H$_{tot}$, falls below 0.01. We include thermal, radiation, and turbulent pressure in our models with the turbulent velocity set to 5.0 km s$^{-1}$.  
\par
In the left panel of Figure~\ref{fig:bpass}, we plot the predicted \ion{He}{1} 1.868$\micron$/Pa$\alpha$ intensity ratios vs. stellar age for the resulting model, along with the observed line ratios from the various apertures indicated by the horizontal lines. The \ion{He}{1} 1.868$\micron$ line is one of the strongest \ion{He}{1} lines in the spectrum and close in wavelength to the Pa$\alpha$ line, ensuring that the line ratio will not be affected by extinction. The predicted \ion{He}{1} 1.868$\micron$/Pa$\alpha$  flux ratios decreases with stellar age as the youngest stars evolve off the main sequence.  As can be seen from the figure, the observed \ion{He}{1} 1.868$\micron$/Pa$\alpha$ in all apertures is consistent with a young stellar population, with little variation in the stellar age. In the right panel of Figure~\ref{fig:bpass}, we plot the predicted \ion{He}{2} 1.863$\micron$/Pa$\alpha$ flux ratio, with the upper limit of the ratio for the nuclear aperture shown by the dotted horizontal line. The infrared \ion{He}{2} lines are much weaker than the optical and UV \ion{He}{2} lines. The lack of detection of this line in a stellar population is not unusual. As can be seen, the upper limit of the observed flux ratio in J1201+0211 is two orders of magnitude above that predicted for the youngest stellar populations. 
\par
Given the \ion{He}{1}/Pa$\alpha$ flux ratio and the ionizing photon flux observed in the nuclear aperture, the stellar cluster according to the photoionization modeling conducted would correspond to a $\leq$ 5~Myr population with a total stellar mass of $\sim3\times 10^4$\,M$_\sun$, somewhat lower than the lowest mass NSCs recently studied by \citet{2022A&A...667A.101F} in more massive nucleated late-type dwarfs. The total luminosity of the cluster if we assume an age of $\sim$ 3Myr would be $\sim3\times 10^{41} $\,erg\,s$^{-1}$. While the NSC scenario is compatible with the JWST spectrum, the steeply rising continuum is impossible to replicate from dust heated by stars \citep{satyapal2018}. It is clear that star formation in low metallicity galaxies can heat the dust to high temperatures in a rare population of dwarf galaxies \citep{2011ApJ...736L..22G, 2011A&A...536L...7I}, but these few cases are characterized by significantly redder $W2-W3$ colors than observed in J1201+0211. While starbursts can mimic obscured AGNs in their red $W1-W2$ colors \citep{2016ApJ...832..119H, satyapal2018}, \citet{satyapal2018} show that it is impossible to replicate the $W2-W3$ colors of J1201+0211 with dust heated by  a purely star-forming low metallicity stellar population. Importantly, there is no indication of a hardening of the radiation field in the nuclear source compared with any other region in the FOV, which is at odds with the hypothesis that an unusually hard radiation field visible at near-infrared wavelengths in the central source is heating the dust to high temperatures causing the steeply rising continuum in the unresolved nuclear source.
\par
If there exists a young NSC in J1201+0211 with age $\leq$ 5~Myr, we can estimate the X-ray luminosity expected from such a NSC. For the first $\sim$ 100-300 Myr, high mass X-ray binaries (HMXBs) dominate the X-ray power output of a stellar population. Using the theoretical binary population synthesis models from \citep{fragos2013} for an instantaneous burst with metallicity $Z=0.1Z_\sun$, the peak expected X-ray luminosity from a young stellar population  $\leq$ 5~Myr is $\sim 6\times10^{37}$\,erg\,s$^{-1}$, about an order of magnitude lower than the observed upper limit for the hard X-ray luminosity. The X-ray non-detection in our XMM-Newton observations is therefore compatible with the NSC scenario for the central source.
\input{panels/bpass_panels}
\par
We note that there are degeneracies between stellar age and ionization parameter, and a likely non-linear dust to gas ratio with decreasing metallicity \citep{2021A&A...649A..18G}. The purpose of the analysis here is to demonstrate that a simple stellar model can easily replicate the observed properties of J1201+0211 without invoking a harder radiation field.

\subsubsection{An Accreting IMBH?}

The lack of high ionization lines in the JWST spectrum is not expected for an AGN that is unobscured at near-infrared wavelengths. We explore the predicted near-infrared emission line spectrum for a photoionization model with an accreting black hole as a function of black hole mass to compare to the upper limits for the high ionization lines obtained from the J1201+0211 spectrum extracted from the nuclear aperture. Here we use the built in \textsc{xspec} \citep{arnaud1996} function \textsc{agnsed} model from \citet{2018MNRAS.480.1247K}, which models the accretion of the black hole with an outer color-temperature corrected blackbody disc, an inner warm and optically thick Comptonizing region to produce the soft X-ray excess and a hot optically thin corona. We adopt the same geometry, equation of state, and abundances as in our stellar model. To explore the effect of black hole mass, we fix the ionization parameter to $U$ $=10^{-2.0}$, and explore three different black hole masses: $\log{(M_{\mathrm{bh}} / \mathrm{M}_{\odot})} = 3, 5, $ and $8$, as well as fix the Eddington ratio to $10^{-1}$ and the spin to 0. The calculation was stopped at the Hydrogen ionization front as in the stellar model. In Figure~\ref{fig:agn_model}, we show the \ion{He}{2}1.863/Pa$\alpha$ and [\ion{Si}{6}] 1.962~$\micron$ /Pa$\alpha$ flux ratio verses black hole mass for our AGN model. The observed upper limits for J1201+0211 are indicated by the dotted horizontal lines. As can be seen, because the AGN models produce a harder radiation field than the stellar models, there is an enhancement of the \ion{He}{2}1.863/Pa$\alpha$  and [\ion{Si}{6}] 1.962~$\micron$  /Pa$\alpha$ flux ratios, an effect that is more  enhanced for lower black hole masses, due to the shift in the SED into the soft X-rays with decreasing black hole mass, an effect previously explored in \citet{Cann2018}. As can be seen, the predicted line ratios are all well above the measured upper limits in J1201+0211. 
\par
These models demonstrate that if a moderately accreting IMBH exists in J1201+0211, it should have been detected, \textit{unless it is deeply embedded and obscured even in the near-infrared}. We again note that there are many assumptions in the model, and caveats that have not been explored. For example, coronal line emission can be heavily suppressed in dusty gas due to depletion onto dust grains (McKaig et al. 2024, in press). The purpose here is to demonstrate that according to this simplified model, high ionization lines should have been detected given the sensitivity of the JWST spectrum, if J1201+0211 harbors an AGN that is unobscured at near-infrared wavelengths.

\input{panels/MBHmodels}
\par

\subsubsection{Where are the missing IMBHs?}

As noted above, our JWST NIRSpec observations cannot rule out the possibility that a deeply embedded AGN exists in the central NSC that remains invisible even at near-infrared wavelengths. The lack of an X-ray detection, similarly does not rule out the possibility of an accreting IMBH. A growing number of AGNs in low mass galaxies are being found to be X-ray weak \citep{2016A&A...596A..64S, 2020ApJ...895..147C, 2020ApJ...894L...5B}, and recent work on the high-z broad line AGNs and ''little red dots'' discovered by JWST has revealed that despite their optical signatures of accretion activity, the vast majority are undetected in the X-rays \citep{2024ApJ...969L..18A,2024arXiv240303872J,2024arXiv240500504M, 2024arXiv240413290Y}, a result that seems to also apply to high-z narrow line AGNs discovered by JWST \citep{2023arXiv231118731S,2024arXiv240218643C}. The lack of X-ray emission and high ionization lines in the near-infrared spectrum therefore does not rule out the accreting IMBH scenario in the central NSC. NSC and SMBHs are known to co-exist in many galaxies \citep[e.g.,][]{2003ApJ...588L..13F, 2008ApJ...678..116S, 2012AdAst2012E..15N, 2019ApJ...872..104N}. If there is simply a young NSC in J1201+0211, there is no explanation for the steeply rising continuum in the unresolved nuclear source. As stated above the uniform \ion{He}{1}/Pa$\alpha$ ratio indicates that there is no hardening of the radiation field in the nuclear source compared to the surrounding, making it difficult to understand why the rising continuum is only seen in the central unresolved source.  
\par
\begin{figure*}
\centering
\includegraphics[width=0.95\textwidth]{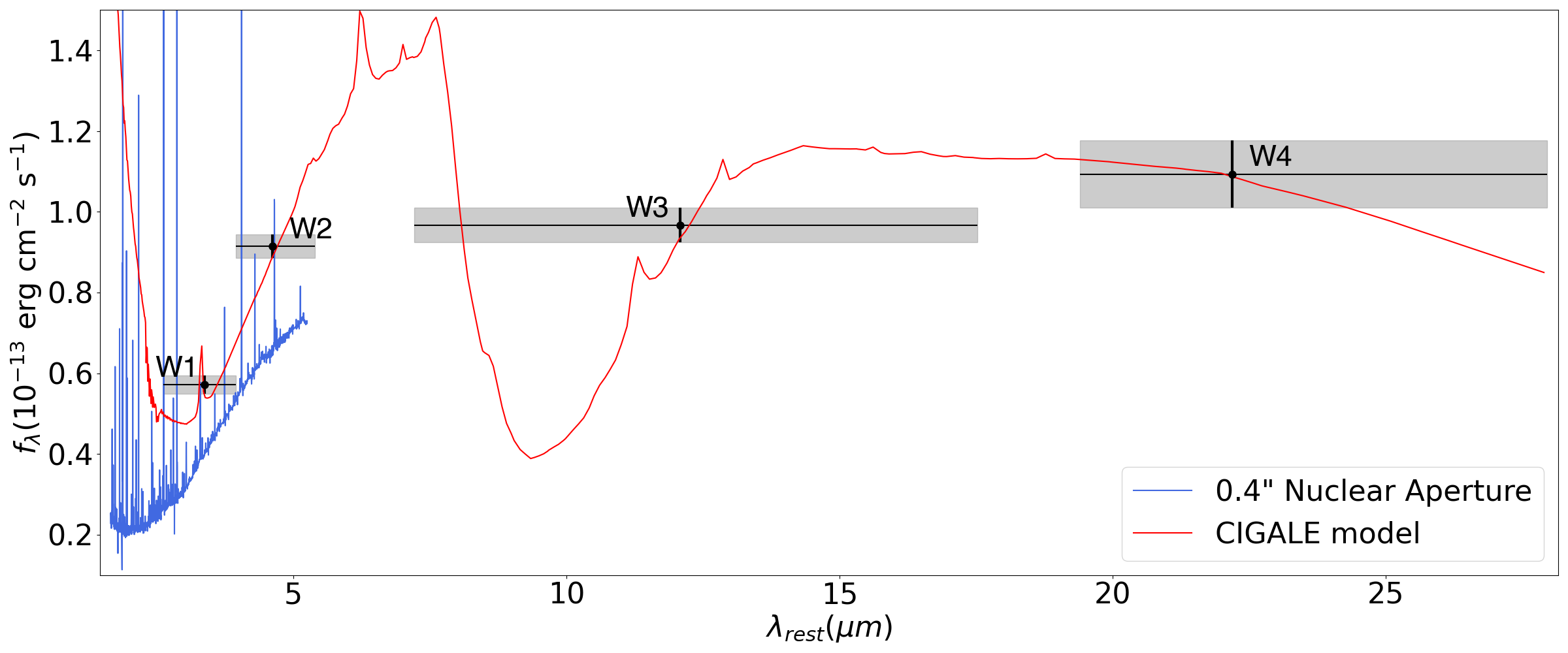}\\
\caption{ JWST NIRSpec spectrum of the nuclear aperture (aperture radius = 0.4\arcsec) with the WISE photometry overlaid. The WISE photometry is fit with a CIGALE model of an obscured AGN, which is overplotted in red. As can be seen, the WISE photometry is well explained by a highly obscured AGN model. The AGN luminosity is $\sim$ $2\times10^{41}$\,erg\,s$^{-1}$, implying a minimum black hole mass of $\sim$ 1450\,M$_\sun$, based on the Eddington limit, roughly consistent with that expected based on extrapolations of black hole galaxy scaling relations derived for more massive black holes . }
\label{fig1}
 \label{fig:sed_nucleus}
\end{figure*}

To explore the possibility that a deeply embedded AGN is responsible for the red WISE source, which we have demonstrated originates in an unresolved central source with steeply rising continuum, we carried out a fit to the WISE photometry using the CIGALE code \citep{2022ApJ...927..192Y}. In Figure~\ref{fig:sed_nucleus}, we show the nuclear aperture spectrum with the WISE photometry overlaid. We use the SKIRTOR models for the AGN continuum \citep{2016MNRAS.458.2288S}, which include accretion disk, AGN torus, and polar dust emission. For the model fit, we fix the extinction to the value required to suppress the near-infrared coronal line ($\tau _{2\micron}$ = 3.39; see section 5.4 and Figure 5) and adopt the SMC extinction law choice in CIGALE for the dust emission. The resulting best fit to the WISE photometry for this embedded AGN scenario is shown in Figure~\ref{fig:sed_nucleus} together with the JWST spectrum for the nuclear aperture. The resulting bolometric luminosity of the embedded AGN from the best fit CIGALE model is $\sim$ $2\times10^{41}$\,erg\,s$^{-1}$. The luminosity of the purported embedded AGN would imply a minimum black hole mass of  $\sim$ 1450\,M$_\sun$, based on the Eddington limit, roughly consistent with that expected based on extrapolations of black hole galaxy scaling relations derived for more massive black holes \citep{2013ARA&A..51..511K}. If this scenario is true, a deeply embedded AGN obscured even at near-infrared wavelengths inside a young NSC hints at the possibility of a stellar cluster origin for the black hole, and a possible evolutionary connection between SMBHs and NSC.

Deeply embedded AGNs that remain hidden at near-infrared wavelengths are not unprecedented. Indeed recent JWST NIRSpec observations of Arp 220 show that there is no evidence of the purported AGN in this source in the near-infrared \citep{2024arXiv240313948P}. While a deeply embedded AGN in a ULIRG is not surprising, the possibility of an embedded AGN in a metal poor dwarf is unusual. Indeed, Herschel data from the Dwarf Galaxy Survey \citep{2013PASP..125..600M}, indicate that at metallicities comparable to that of J1201+0211, the dependence of dust-to-gas ratio is steeper than the linear relation found for higher metallicity galaxies, implying relatively less dust (or more gas) at lower abundances \citep{2014A&A...563A..31R}. However, extreme dust content is not unprecedented in XMP dwarfs, as is seen in the young star forming dwarf SBS 0355-052E \citep{2016MNRAS.457.1842S}, in which extreme red WISE colors, MIR variability and a [\ion{Ne}{5}] line have been detected \citep{2023arXiv230403726H}. The AGN bolometric luminosity of the purported AGN in SBS 0355-052E estimated from its SED is about $\sim$ 70 times larger than that estimated for the purported embedded AGN in J1201+0211. The black hole mass based on the broad H$\alpha$ line in this source is almost two orders of magnitude larger than J1201+0211, potentially making J1201+0211 a galaxy that harbors the lowest mass IMBH in a dwarf galaxy that has been found so far. Our observations have revealed that there is clear evidence for dust in the galaxy, although the near-infrared derived extinction would not be sampling the extinction toward a  deeply embedded AGN, should one exist. Apart from attenuation, the presence of dust can have significant effects on the line emission from ionized gas. In recent work, McKaig et al. (in press) find that the presence of dust has a dramatic effect on the coronal line emission, suppressing the line luminosity by as much as 3 orders of magnitude compared with dust-free gas, primarily because of gas-phase depletion onto dust grains. Highly refractory elements such as Silicon and Calcium are more severely affected by the presence of dust than the noble gases, such as Neon. Mid-infrared observations with MIRI MRS are crucial to reveal a possibly hidden IMBH in this source. The [\ion{Ne}{5}] 14~$\micron$ line is one of the strongest lines detected in AGNs and has been shown to uncover AGNs not identified through any other means in late-type lower mass galaxies \citep{2007ApJ...663L...9S, 2008ApJ...677..926S, 2009ApJ...704..439S}. Longer wavelength observations of J1201+0211 are the crucial next step in confirming or refuting the presence of an embedded AGN in this source.

\section{Conclusions}

We present JWST NIRSpec IFS observations of the extremely metal poor local dwarf galaxy, J1201+0211. These observations represent the first JWST NIRSpec observations of an XMP dwarf. We also obtained XMM-Newton observations of this source. Our main results can be summarized as follows:

\begin{enumerate}

\item The unprecedented sensitivity of JWST reveals a remarkably rich spectrum with more than 100 emission lines in the 1.7-5.2$\micron$ region. As a reference, we provide a list of all identified features in an aperture centered on the nucleus for this primordial galaxy analog.

\item We detect a bright unresolved central continuum source with a steep spectrum well-fit by a power law with index consistent with hot dust in AGNs ($F_\nu \approx\nu^{-1.5}$). This steeply rising continuum is only found in the central unresolved source; all other continuum sources are characterized by a relatively flat spectrum, consistent with star forming regions. These observations reveal that the red source identified by WISE is extremely compact and centrally located, with a spatial extent $\lessapprox $ 3 pc.

\item The spectrum within the IFS field of view for this metal poor dwarf is dominated almost exclusively by hydrogen recombination lines and ro-vibrational lines from molecular hydrogen, with a few He I lines. There are no He II lines, no PAH features, no CO absorption features typically seen in late-type stars, no ice features, and no evidence for any coronal lines in the spectrum. 

\item Intricate structure on parsec scales is seen in the ionized gas, with a bright central source, an ionized cavity with a shell-like structure often seen in late-type dwarfs. The warm molecular gas and the ionized gas show different morphologies, with variations seen in $H_{2}$ line ratios indicating variations in the excitation of the molecular gas across the FOV. The hydrogen recombination line ratios indicate patchy extinction toward the near-infrared emitting ionized gas, with regions with relatively high extinctions ($A_{V} \sim$ 8 mag) for a metal poor dwarf.   

\item We compared the JWST spectrum in the nuclear aperture to that predicted from a stellar population. We find that the near infrared emission line spectrum can be explained by a young stellar population ( $<$ 5~Myr), with little variation in the stellar age throughout the central $\sim$ 200~pc region. The ionizing photon flux of the nuclear source is Q(H) $\sim  3\times10^{51}$\,s$^{-1}$, comparable to the total ionizing flux from the giant star cluster R136 in 30 Doradus in the LMC \citep{1998MNRAS.296..622C}.  Given the stellar age, the NSC mass would be $\sim3\times 10^4$\,M$_\sun$ .

\item We do not detect an X-ray source in J1201+0211. The upper limit to the luminosity in the hard band is $L_{2-10\,\mathrm{keV}}$ $\sim 5\times10^{38}$\,erg\,s$^{-1}$, above that expected for the modeled stellar population.

\item Although we do not detect any X-ray emission or high ionization lines in this metal poor dwarf, the presence of a young NSC alone does not easily explain the steeply rising infrared SED in the unresolved nuclear source. There is no evidence for a hardening of the interstellar radiation field in the nucleus compared to surrounding regions indicating that the dust is heated to higher temperatures by the stellar population. The dust properties or geometry in the central cluster may somehow give rise to an unusually red MIR continuum source.

\item Based on our observations, we cannot rule out the possibility of a deeply embedded accreting IMBH in the nuclear source. If we assume the steeply rising continuum in the unresolved nuclear source is due to a deeply embedded AGN, the bolometric luminosity of the embedded AGN from an SED fit to the WISE photometry is $\sim$ $2\times10^{41}$\,erg\,s$^{-1}$. The luminosity of the purported embedded AGN would imply a minimum black hole mass of  $\sim$ 1450\,M$_\sun$, based on the Eddington limit, roughly consistent with that expected based on extrapolations of black hole galaxy scaling relations derived for more massive black holes. Observations with MIRI MRS are crucial to determine if a hidden IMBH resides in this source

\item These observations reveal that either a metal poor stellar population can heat the dust to extremely high temperatures, a result that would have significant impact on the reliability of MIR color selection in AGN surveys and our understanding of metal poor stars, or an accreting IMBH is deeply embedded and hidden even at near-infrared wavelengths.

.

\end{enumerate}

\section{Acknowledgements}

The authors are very grateful for the dedicated community who made the observatory a reality, and for the complex pipeline development work over the course of the year carried out by STScI, the ERS teams, and the community at large. The authors gratefully acknowledge Remington Sexton for the development of the spectral fitting code used in this work, and to Michael Reefe for helping to adapt it to IFS observations. 

This work is based on observations made with the NASA/ESA/CSA JWST. The data were obtained from the Mikulski Archive for Space Telescopes at the Space Telescope Science Institute, which is operated by the Association of Universities for Research in Astronomy, Inc., under NASA contract NAS 5-03127 for JWST. STScI is operated by the Association of Universities for Research in Astronomy, Inc., under NASA contract NAS5-26555. Support to MAST for these data is provided by the NASA Office of Space Science via grant NAG5-7584 and by other grants and contracts. These observations are associated with program \#1983. The JWST data used in this paper can be found in doi:10.17909/ct0b-sh28. S. D. and S. S. gratefully acknowledge funding from JWST Cycle 1 grant GO-01983.001-A.

This publication makes use of data products from the Wide-field Infrared Survey Explorer, which is a joint project of the University of California, Los Angeles, and the Jet Propulsion Laboratory/ California Institute of Technology, funded by the National Aeronautics and Space Administration. This research has made use of the NASA/IPAC Infrared Science Archive, which is funded by the National Aeronautics and Space Administration and operated by the California Institute of Technology. 

This work made use of v2.2.1 of the Binary Population and Spectral Synthesis (BPASS) models as described in Eldridge, Stanway et al. (2017) and Stanway \& Eldridge et al. (2018).

This research made use of Astropy,\footnote{\url{http://www.astropy.org}} a community-developed core Python package for Astronomy \citep{2013A&A...558A..33A}, as well as \textsc{topcat} \citep{2005ASPC..347...29T}. The pipeline processing, spectral fitting, and Cloudy simulations carried out in this work were run on ARGO and HOPPER, research computing clusters provided by the Office of Research Computing at George Mason University, VA. (\url{ http://orc.gmu.edu})

\bibliographystyle{yahapj}
\bibliography{main}

\appendix
\input{tables/line_ident_all}
\(\) 

\end{document}

%% file: panels/BPTWISE.tex
\begin{figure*}
    \centering
    \includegraphics[width=\columnwidth]{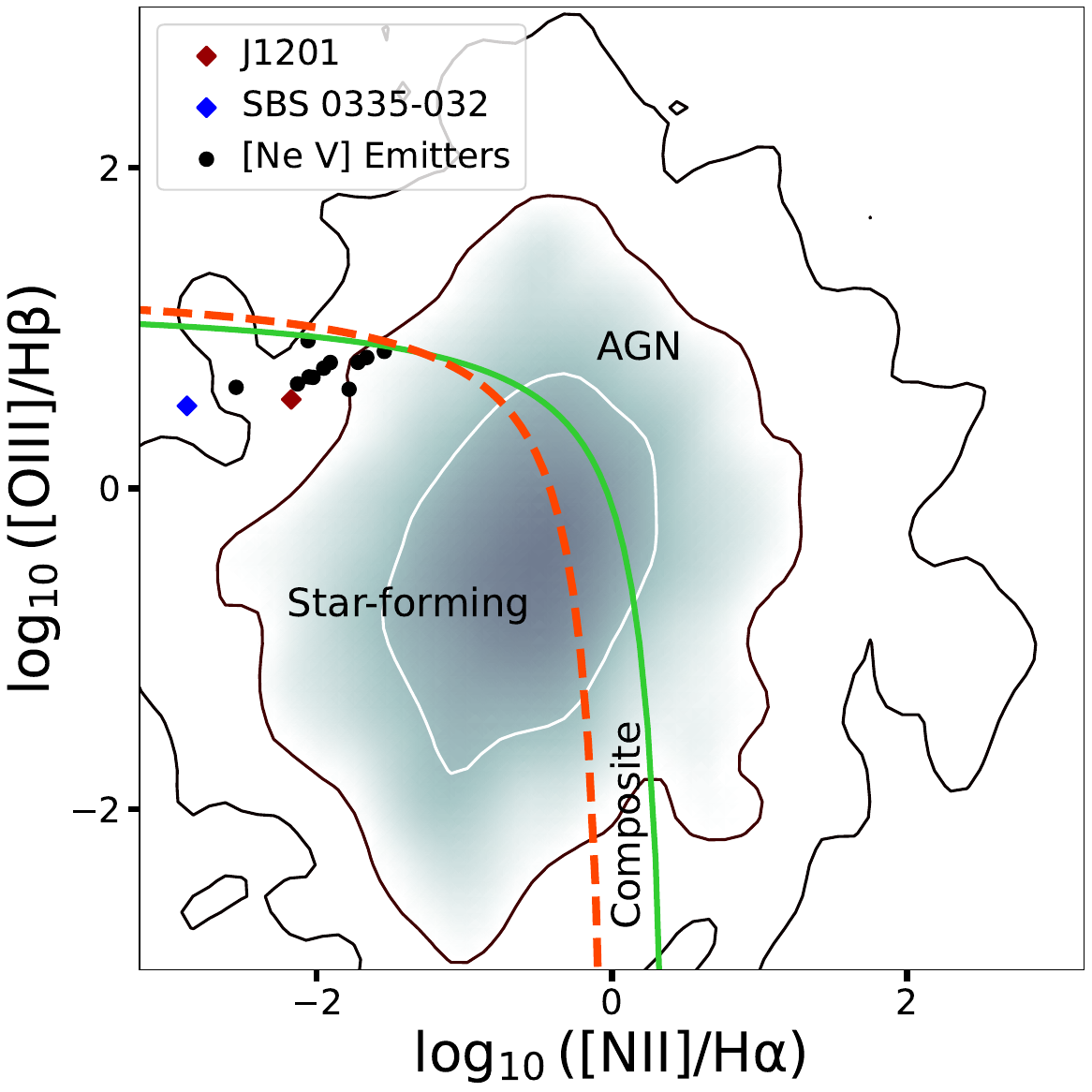}
     \includegraphics[width=\columnwidth]{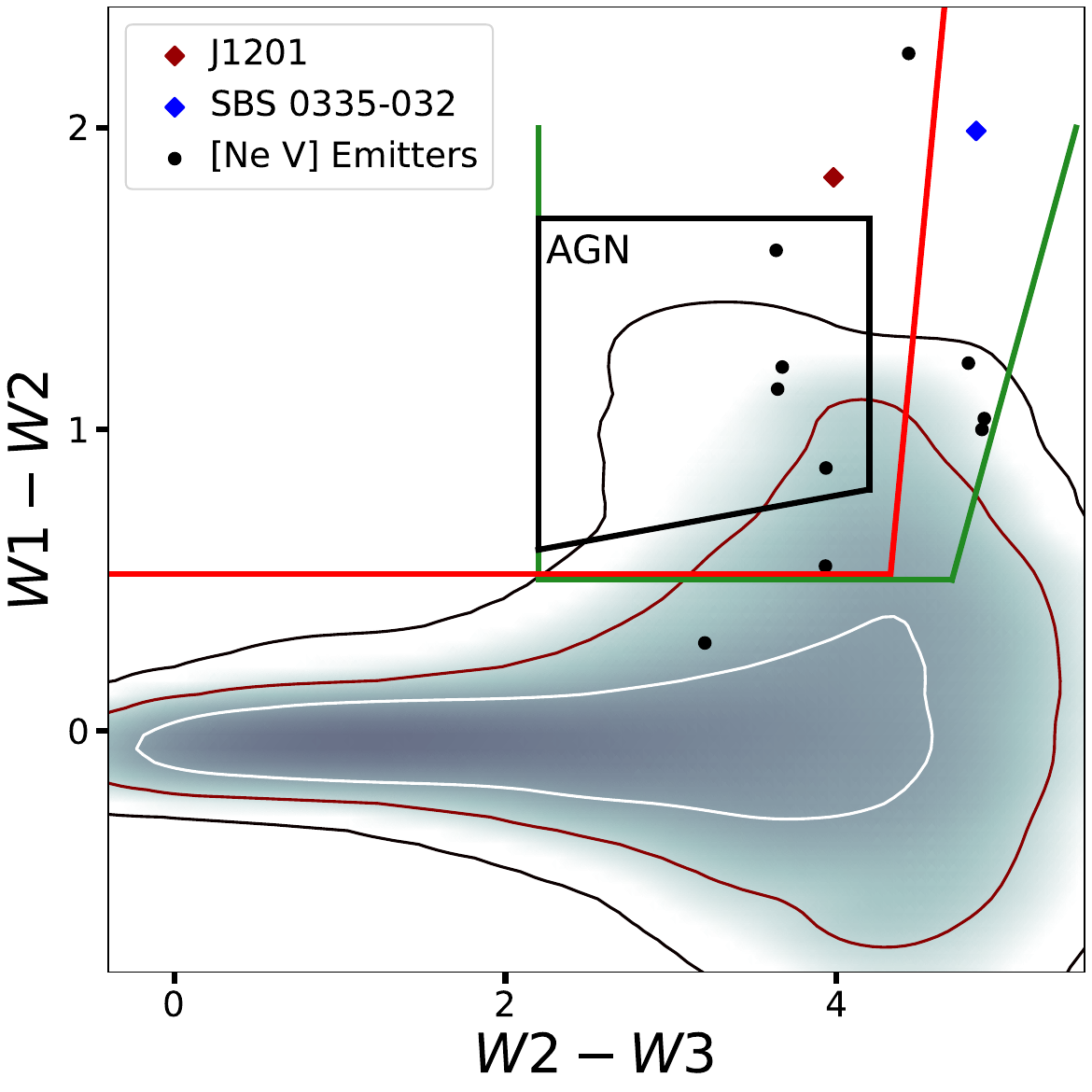}
      
    \caption{\textit{Left}: A BPT ratio plot showing the narrow line ratios of J1201+0211 from SDSS indicated by the burgundy diamond. We also display the narrow line ratios for the sample of BCDs showing [\ion{Ne}{5}] emission from \citep{2012MNRAS.427.1229I,2021MNRAS.508.2556I,2024ApJ...966..170H}, which includes the XMP SBS 0335-032, another metal poor dwarf with extreme MIR colors and MIR variability indicative of an AGN \citep{2023arXiv230403726H}. The star-forming, ``composite,'' and AGN regions are marked with text and separated by the curves defined in \citet{2003MNRAS.346.1055K} (solid green line) and \citet{2001ApJ...556..121K} (dashed red line).  The grey shading displays the BPT values for the entire sample of dwarf galaxies from \citet{2013ApJ...775..116R}.\\
    \textit{Right}: A WISE color-color plot with the value for J1201+0211 displayed by the burgundy diamond. We also show the colors of the [\ion{Ne}{5}] emitting dwarfs as in the BPT plot. The AGN demarcation box from \citet{jarrett2011} is shown by the black line, the demarcation box from \citet{satyapal2018} is shown by the red line, and the demarcation box from \citet{2018MNRAS.478.3056B} is shown in green. Note that the vast majority of dwarfs do not show the extreme MIR colors of J1201+0211; unlike J1201+0211, most of the [\ion{Ne}{5}] emitting dwarfs, including SBS 0355-032 do not meet the \citet{satyapal2018} color cut. Note that while J1201+0211 is just above the AGN demarcation box from \citet{jarrett2011}, \citet{satyapal2018} demonstrated that anything inside the red bounding box cannot be replicated by even the most extreme metal poor stellar populations. }
    \label{fig:bptwise}
\end{figure*}

%% file: tables/cl_upper_limits.tex
\startlongtable
\begin{deluxetable*}{ccc}
\tabletypesize{\footnotesize}
\tablecaption{Upper Limits for High Ionization Lines in Nuclear Aperture}
\tablehead{\colhead{Line} & \colhead{$\lambda_{\rm rest}$} & \colhead{Flux Upper Limit}\\
\colhead{} & \colhead{($\mu m$)} & \colhead{(10$^{-17}$ erg cm$^{-2}$ s$^{-1}$)}}
\decimals
\startdata
He II & 1.8630 & < 0.35\\
\lbrack S VI\rbrack & 1.8870 & < 0.41\\
\lbrack Si XI\rbrack & 1.9340 & < 0.25\\
\lbrack Si VI\rbrack & 1.9620 & < 0.23\\
\lbrack Al IX\rbrack & 2.0440 & < 0.12\\
\lbrack Ca VIII\rbrack & 2.3211 & < 0.10\\
\lbrack Si VII\rbrack & 2.4807 & < 0.05\\
\lbrack Al V\rbrack & 2.9040 & < 0.13\\
\lbrack Ca IV\rbrack & 3.2060 & < 0.08\\
\lbrack Al VI\rbrack & 3.6590 & < 0.08\\
\lbrack Si IX\rbrack & 3.9280 & < 0.15\\
\lbrack Mg IV\rbrack & 4.4870 & < 0.22\\
\lbrack Ar VI\rbrack & 4.5280 & < 0.20\\
\enddata
\label{tab:cl_ident}
\end{deluxetable*}

%% file: panels/flux_maps.tex
\begin{figure*}
\centering
\includegraphics[width=\columnwidth]{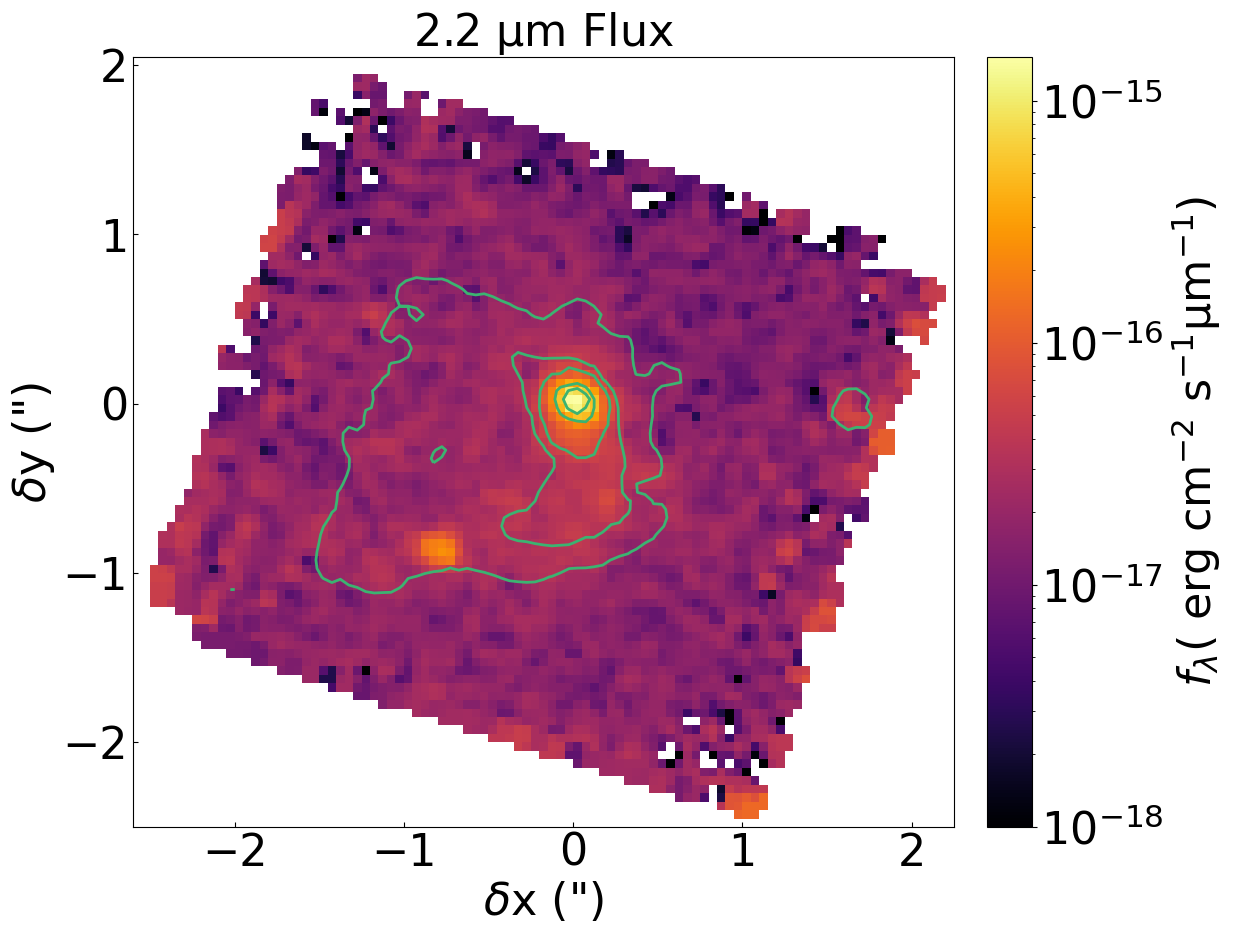}
\includegraphics[width=\columnwidth]{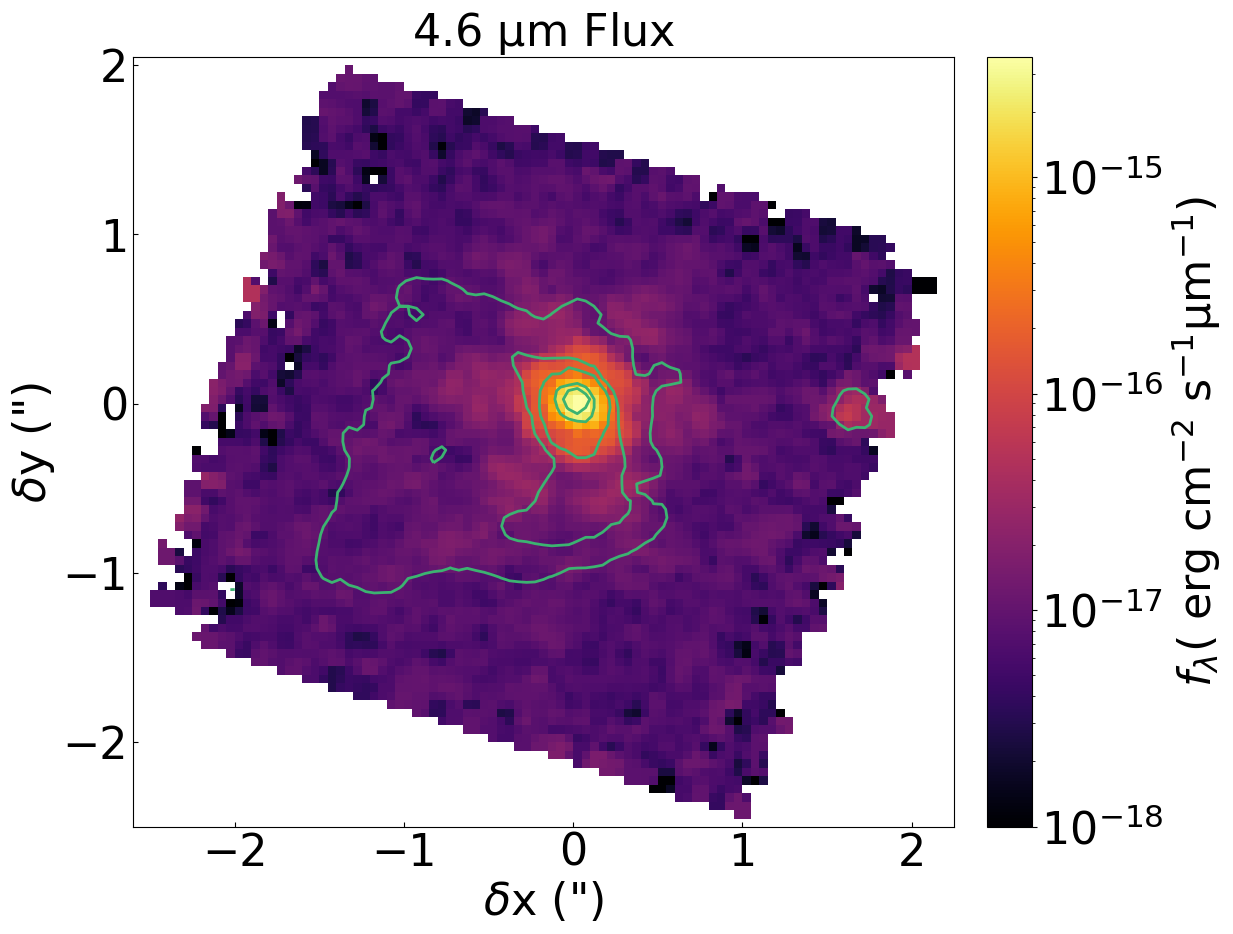}
\includegraphics[width=\columnwidth]{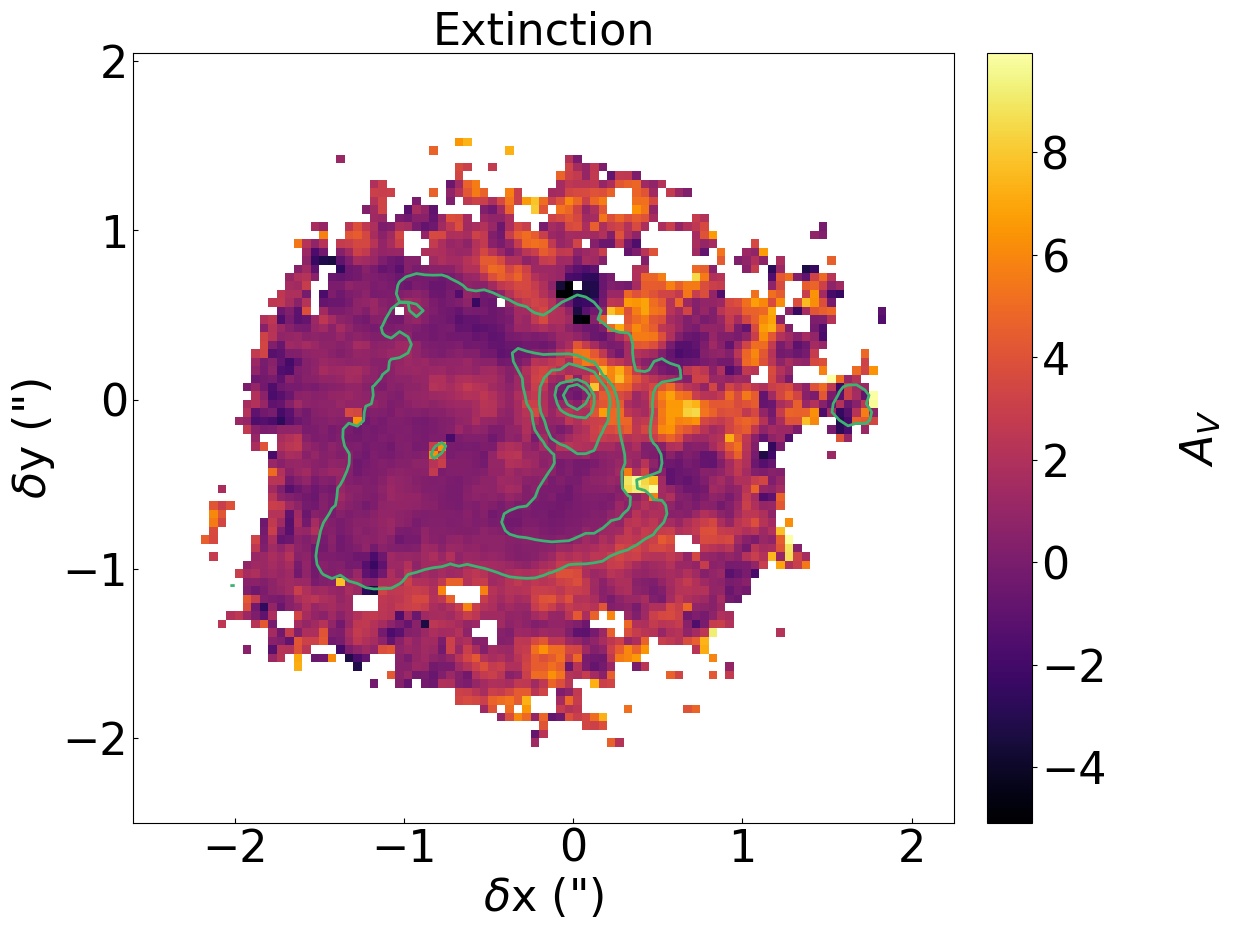}
\includegraphics[width=\columnwidth]{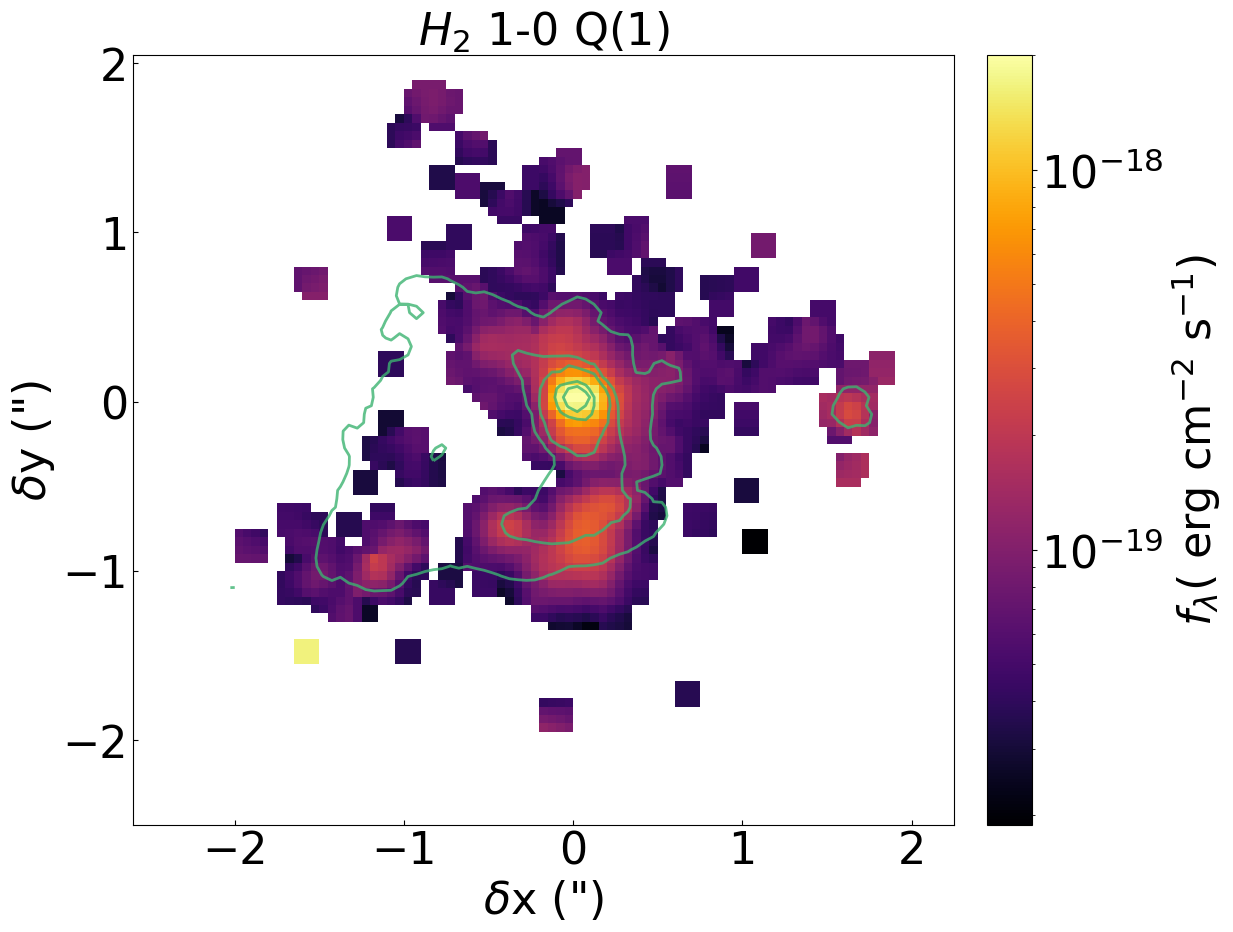}    
\caption{\textit{Top Left}: Map of the line-free continuum at 2.2 $\micron$. The bright central source coincides with the peak in the Pa$\alpha$ image displayed in Figure~\ref{fig:spectra_regions}. \textit{Top Right:} Map of the line-free continuum at 4.6 $\micron$, the center of the W2 band.The southern source disappears in the longer wavelength continuum, and only the central nuclear source appears. Very little diffuse emission is seen in the continuum. The nuclear source is unresolved in the continuum maps, implying a spatial extent of $\lesssim$ 3 pc. \textit{Bottom Left}: The extinction map obtained using the Pa$\alpha$ to Br$\alpha$ ratio (see Section 5.3 for details). The extinction is patchy with the peak extinction seen in the nuclear source. \textit{Bottom Right:}$H_{2}$ 1-0 Q(1) 2.4059 $\micron$ flux map. The morphology of the warm molecular gas maps is distinctly different from that of the ionized gas traced by the hydrogen recombination lines. Note that the flux maps have been masked such that only spaxels with S/N $>3$ are shown. The spaxel size for all maps is 0.05\arcsec; green contours on all maps correspond to the Pa$\alpha$ emission.}
\label{fig:flux_maps}
\end{figure*}

%% file: panels/sample_spectra.tex
\begin{figure*}
    \centering
    \includegraphics[width=\textwidth]{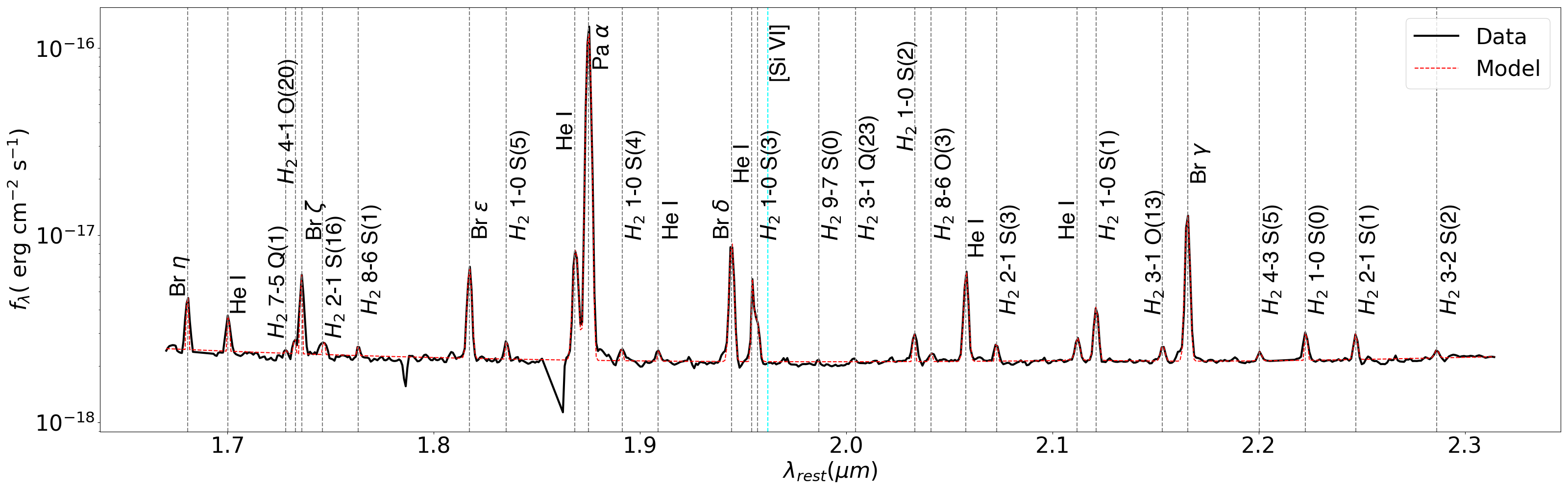}
    \includegraphics[width=\textwidth]{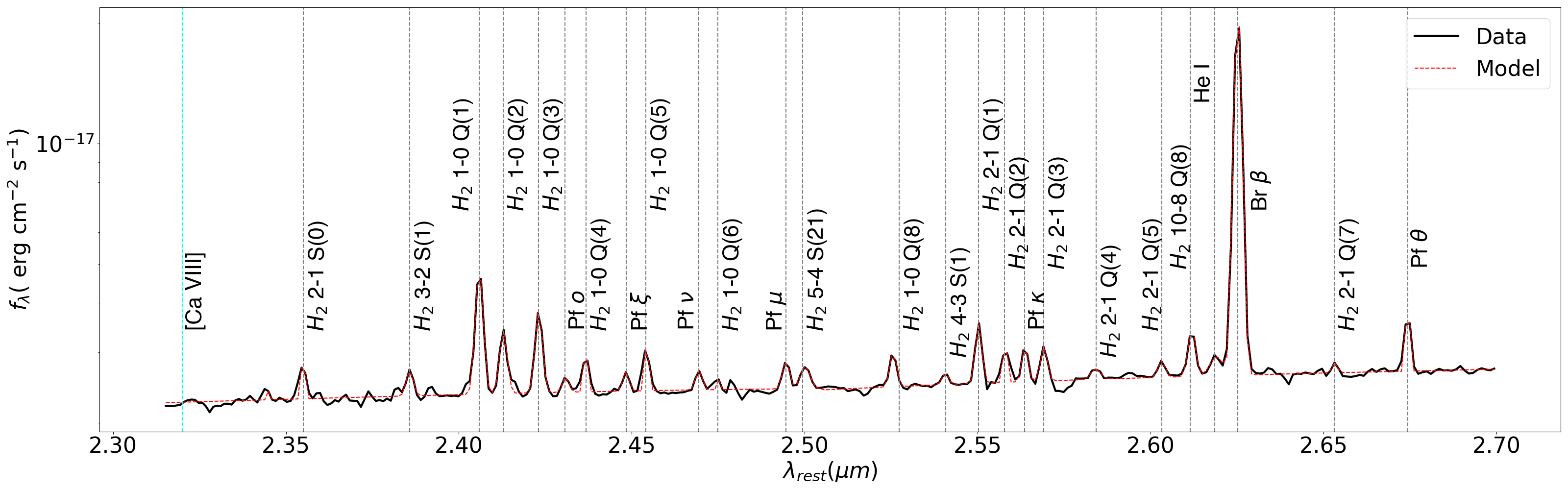}
    \includegraphics[width=\textwidth]{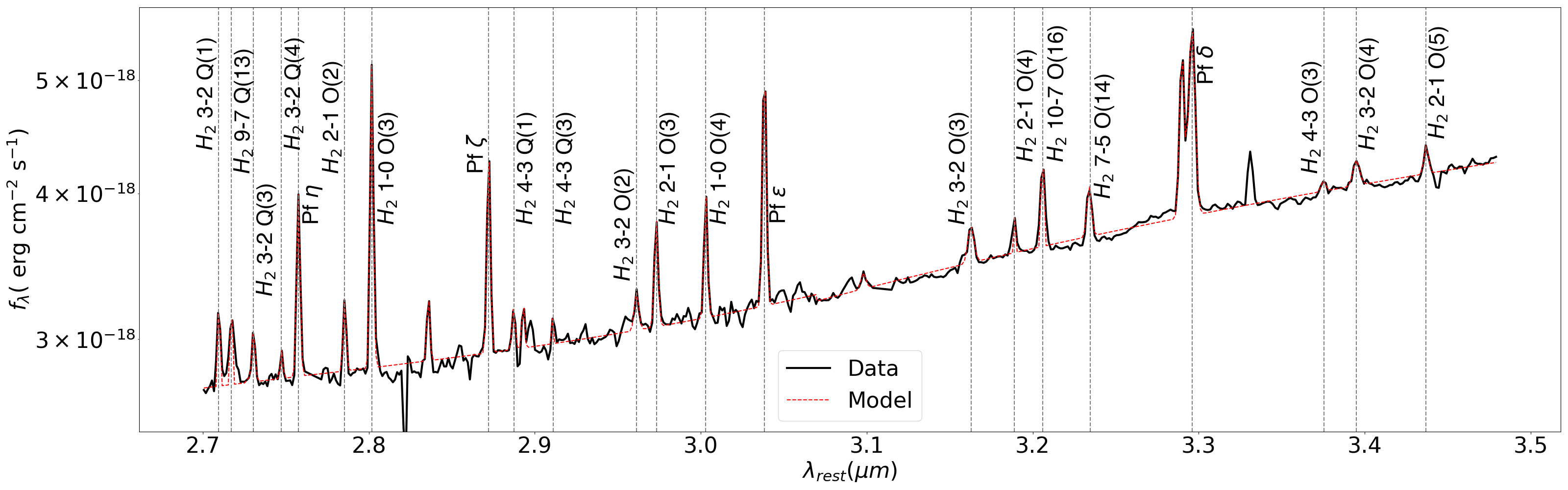}
    \includegraphics[width=\textwidth]{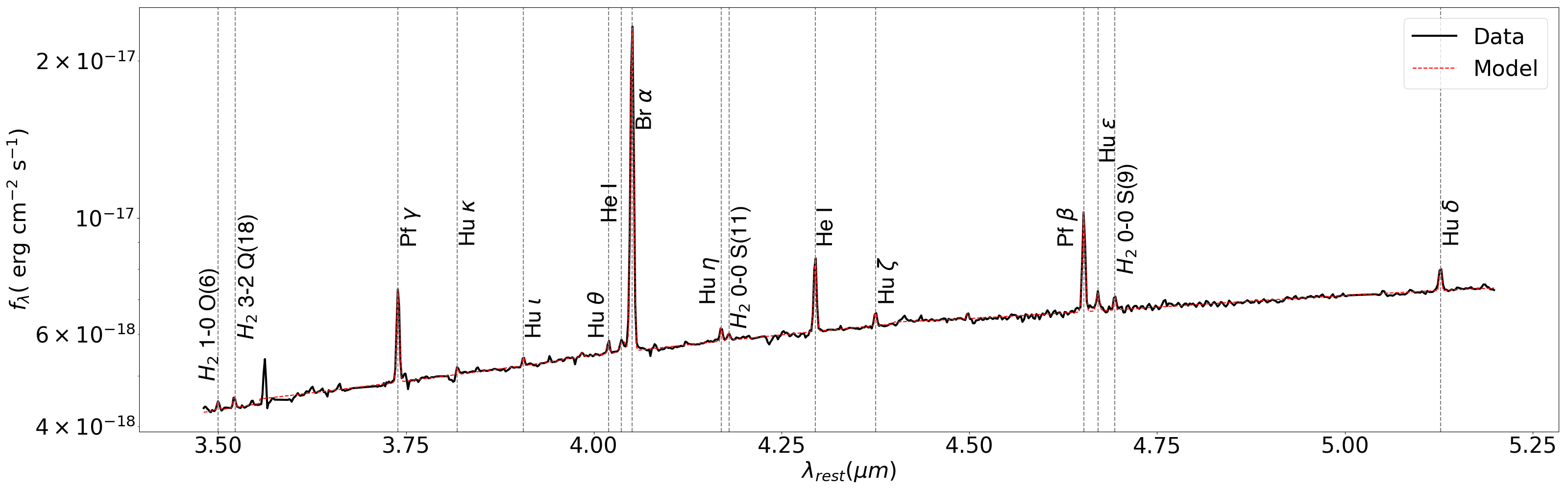}
    \caption{Spectral windows covering the majority of detected lines in the JWST spectrum centered on the nuclear aperture. The JWST data, corrected for redshift, are plotted in black, while the model fits are overlaid on the spectra.  Prominent lines are labeled in each figure.}
    \label{fig:spectra_sample}
\end{figure*}

%% file: panels/coronal_line_ULs.tex
\begin{figure*}
    \centering
    \includegraphics[width=\columnwidth]{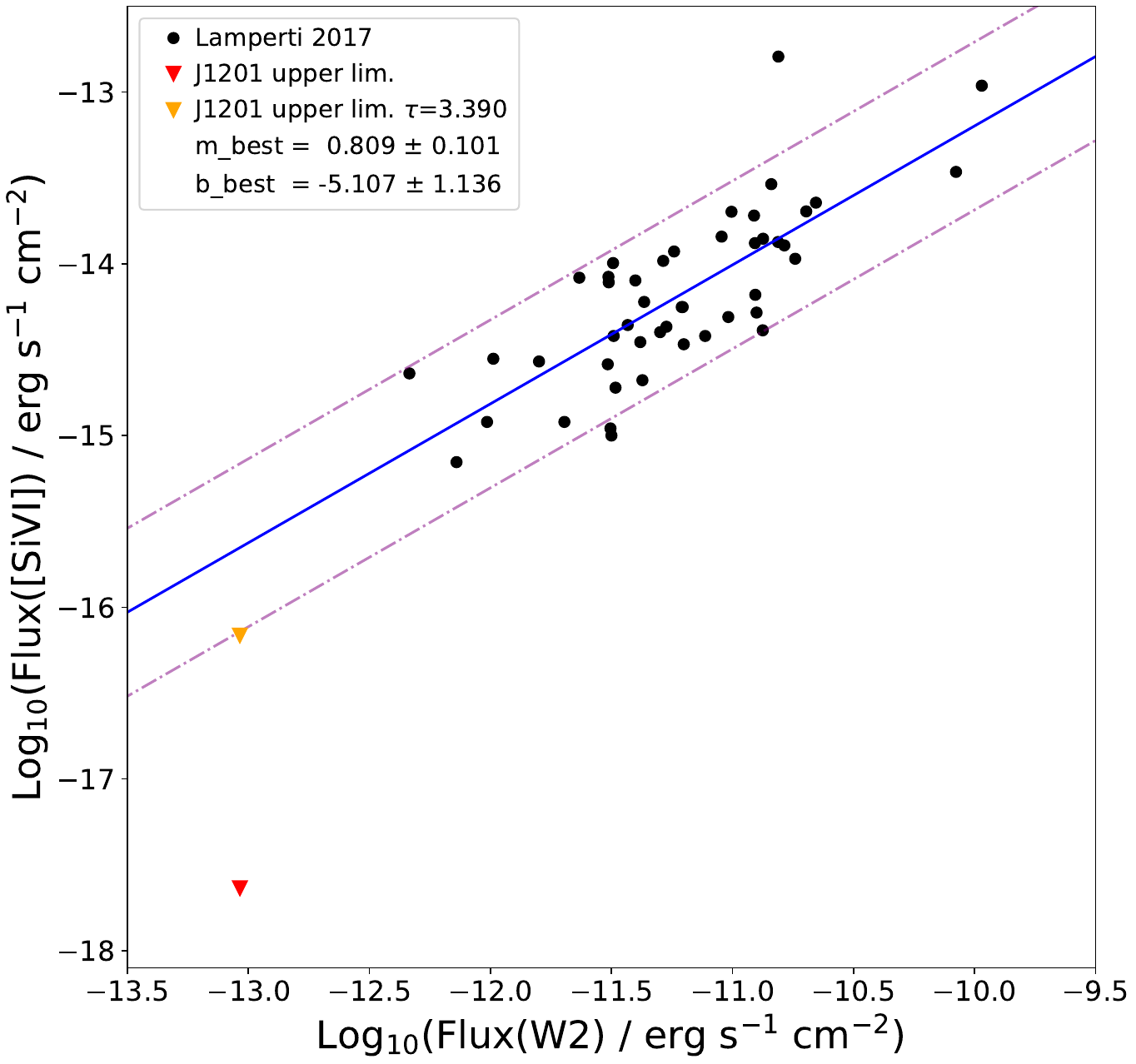}
     \includegraphics[width=\columnwidth]{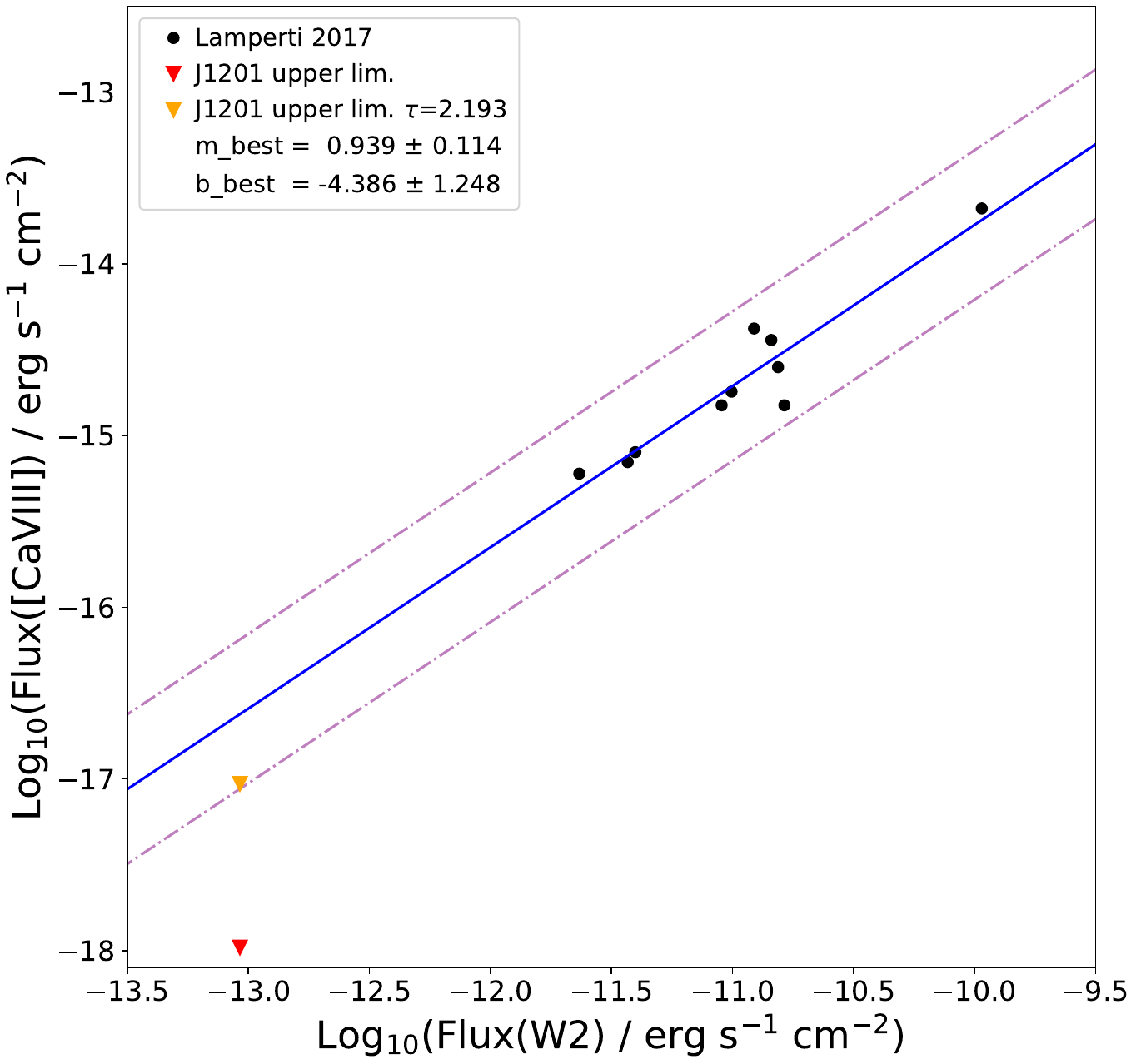}
      
    \caption{\textit{Left:}[\ion{Si}{6}] 1.962~$\micron$ line, the mostly commonly detected near-infrared coronal line in local AGNs,  verses the WISE W2 (4.6~$\micron$) flux for J12011+0211, in comparison with the values for the hard X-ray selected Swift/BAT sample relation from \citet{2017MNRAS.467..540L}. \textit{Right:} Similar plot for the [\ion{Ca}{8}] 2.32$\micron$ line. The solid line in each plot represents the best fit linear regression for the sample of hard X-ray selected AGNs, and the dashed lines represent the 1$\sigma$ scatter in the relation. As can be seen, with the exquisite sensitivity of JWST, the upper limits clearly fall below the extrapolation of the relations based on local AGNs. In each plot, the value of the extinction at the wavelength of each line required to move the upper limit up to the lower bound in the local relations (yellow triangle) is listed in the legend. As can be seen, if an AGN resides in J1201+0211 with similar coronal line properties to local AGNs, it would need to be deeply embedded, invisible at near-infrared wavlengths.}
    \label{fig:CL_ULs}
\end{figure*}

%% file: panels/bpass_panels.tex
\begin{figure*}
    \centering
    \includegraphics[width=\columnwidth]{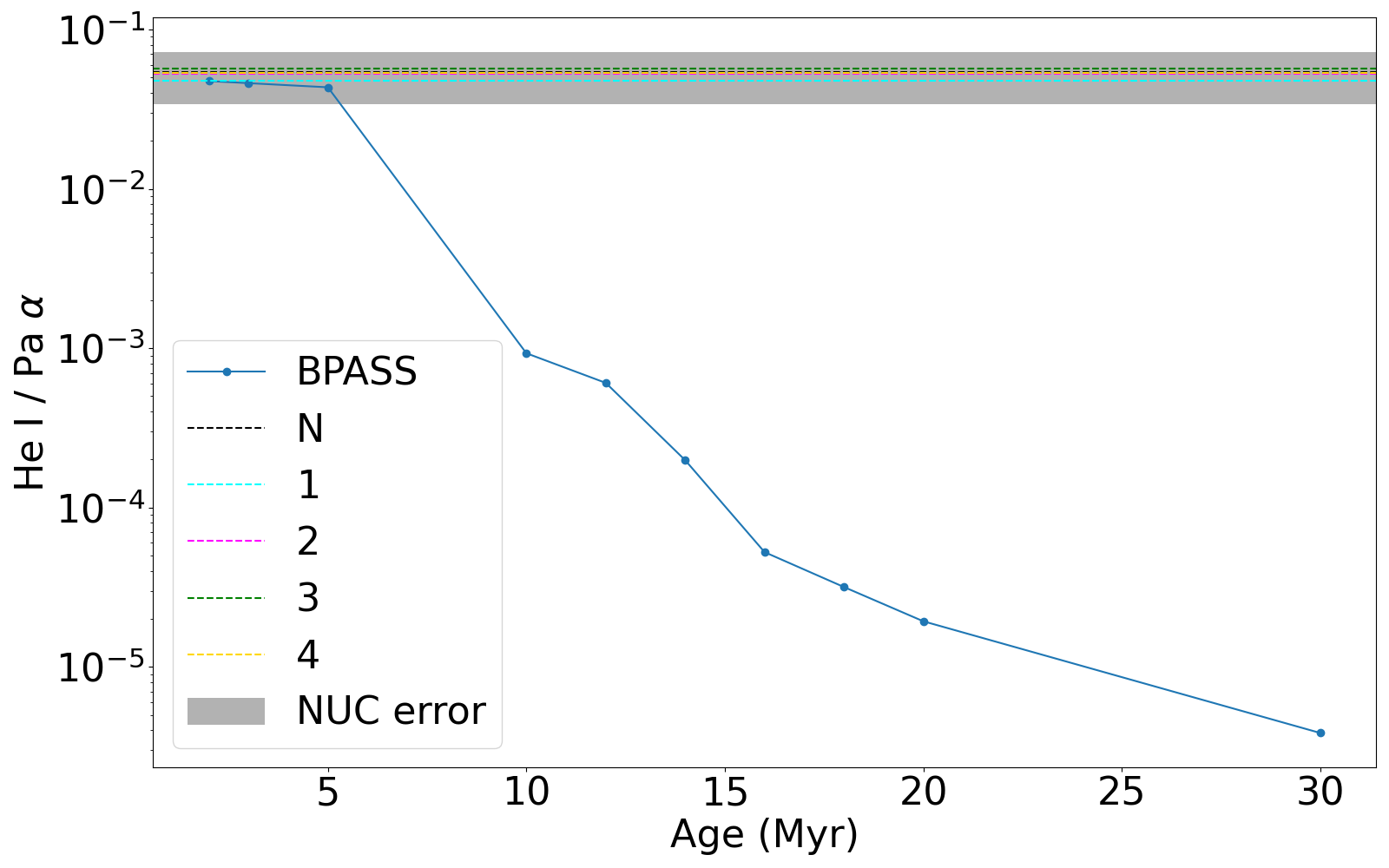}
     \includegraphics[width=\columnwidth]{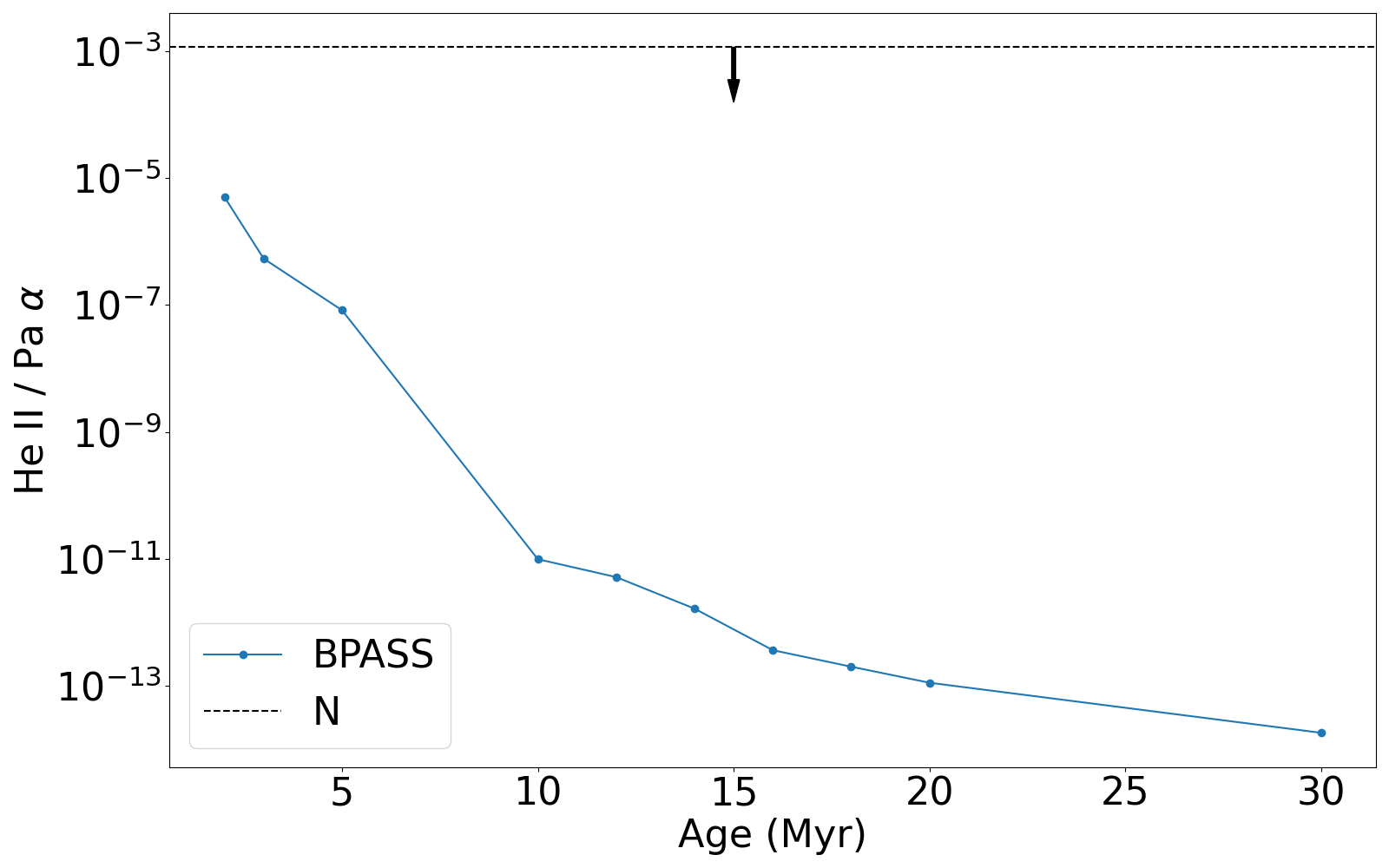}
      
    \caption{\textit{Left:} \ion{He}{1}1.868/Pa$\alpha$ flux ratio vs. stellar age for a single burst BPASS model. Because the ionization potential of $He^{+}$ (24.6~eV) is higher than that of $H^{+}$, the plotted line ratio decreases with age as the stellar population evolves off the main sequence resulting in a softening of the ionizing radiation field. The observed line ratios in the various apertures are indicated by the horizontal lines. The shaded region represents the error in the line ratios at the 95\% confidence interval. As can be seen, the line ratios in all apertures are consistent with a metal poor stellar population, and, importantly, there is no significant difference in the line ratios between the nuclear aperture compared with the other 4 apertures.\textit{Right:} Similar plot for the \ion{He}{2}1.863/Pa$\alpha$ flux ratio. The observed upper limit is shown by the dotted horizontal line. As can be seen, the upper limit for the  \ion{He}{2}1.863 line is consistent with predictions for a young stellar population. }
    \label{fig:bpass}
\end{figure*}

%% file: panels/MBHmodels.tex
\begin{figure*}
    \centering
    \includegraphics[width=\columnwidth]{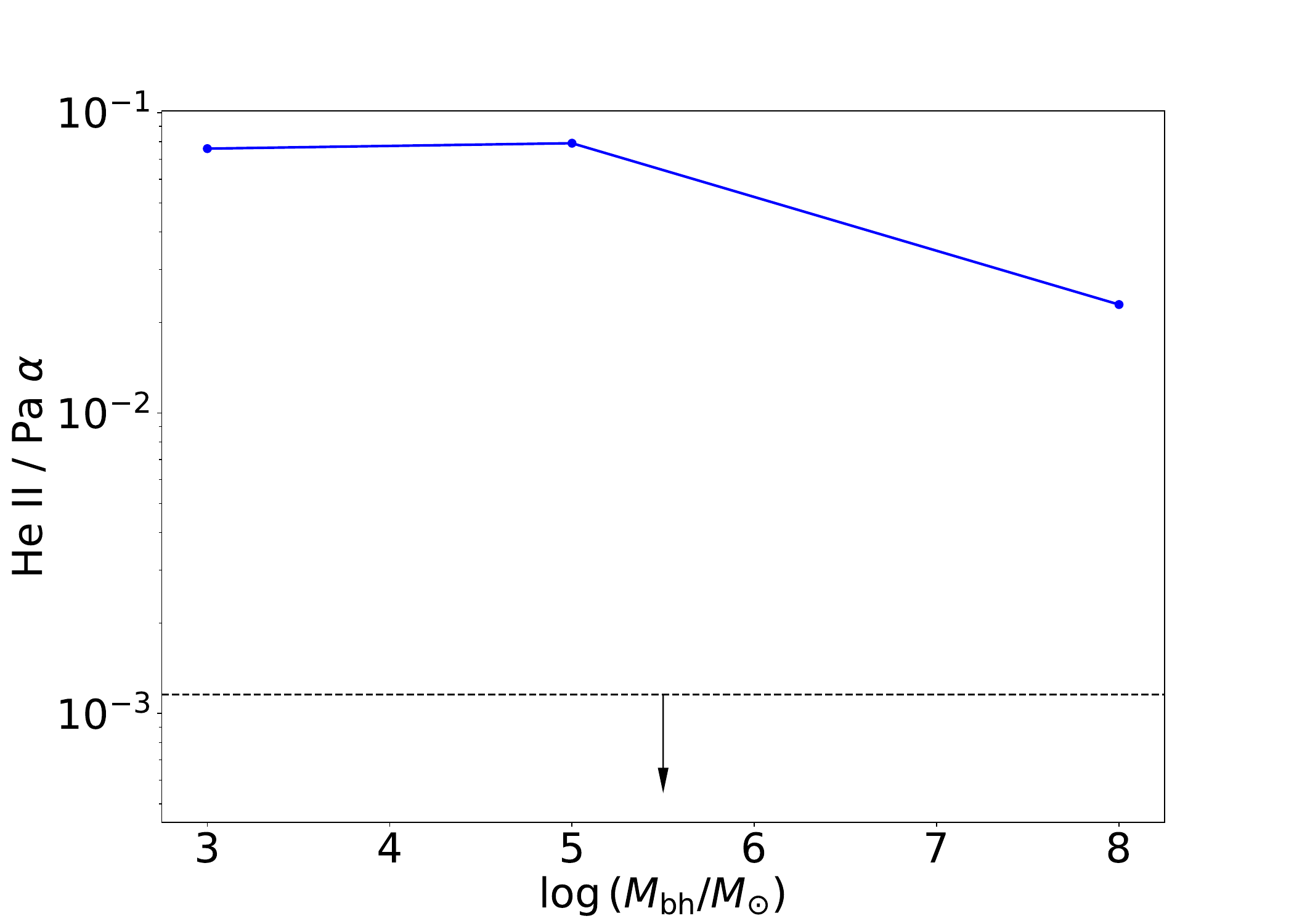}
     \includegraphics[width=\columnwidth]{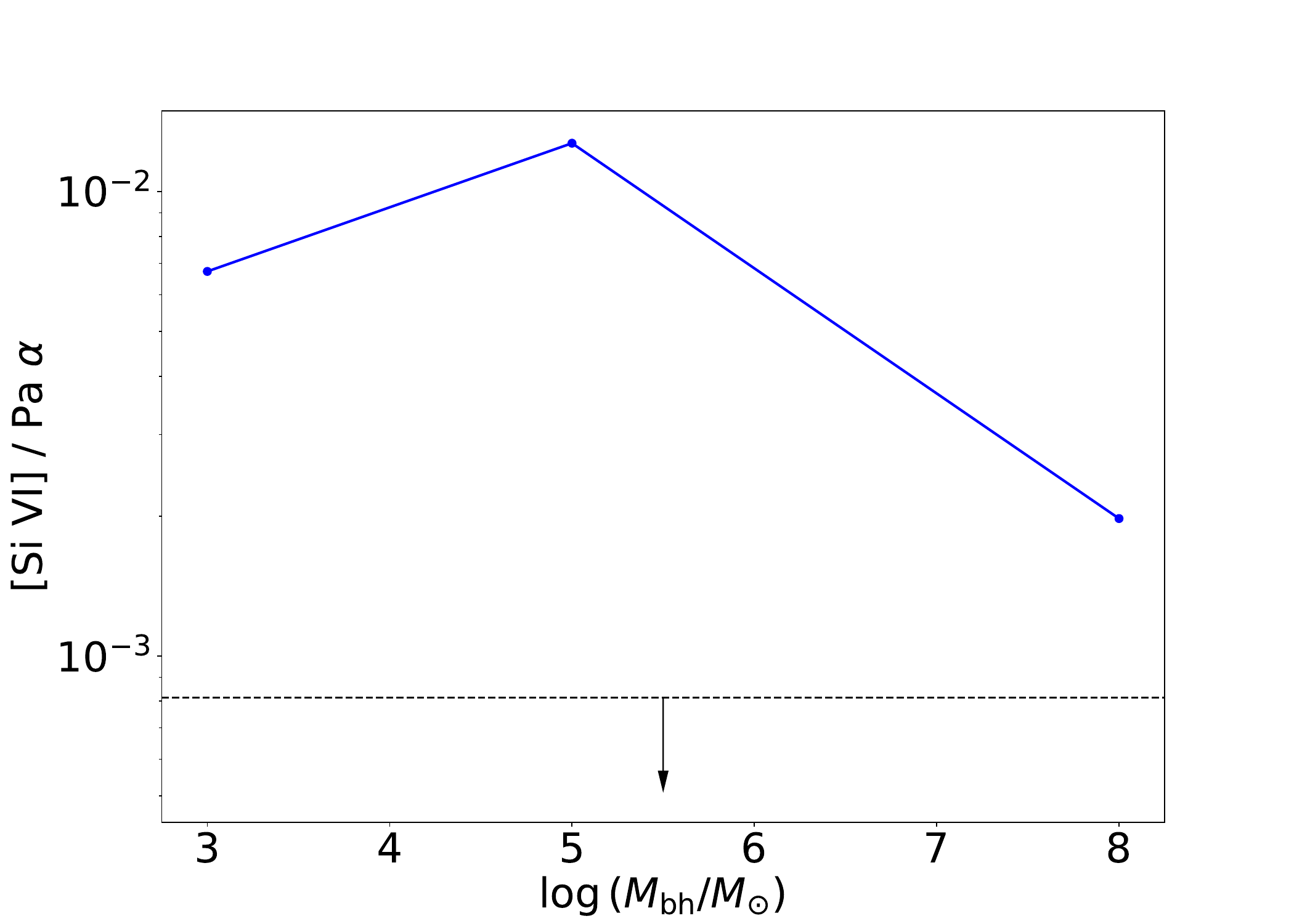}
      
    \caption{\textit{Left:} \ion{He}{2}1.863/Pa$\alpha$ flux ratio vs. black hole mass for a low metallicity AGN model. The observed upper limit is indicated by the dotted horizontal line. Because the AGN models produce a harder radiation field, the ratio of high to lower ionization lines is significantly greater than that predicted by even the youngest metal poor stellar populations. In addition, because the accretion disk temperature increases with stellar mass, the line ratios are enhanced in IMBHs compared with more massive black holes, an effect previously explored by \citet{Cann2018}. Similar plot for the [\ion{Si}{6}] 1.962~$\micron$ /Pa$\alpha$ flux ratio. The observed upper limit is shown by the dotted horizontal line. As can be seen from both plots, the observed upper limit is well below the predictions for all AGN models, indicating that if an AGN resides in J1201+0211, it should have been detected unless it is completely obscured at near-infrared wavelengths.}
    \label{fig:agn_model}
\end{figure*}

%% file: tables/line_ident_all.tex
\startlongtable
\begin{deluxetable*}{ccccc}
\tabletypesize{\footnotesize}
\tablecaption{J1201+0211 Emission Line Fit Parameters for the Nuclear Aperture}
\tablehead{\colhead{Line} & \colhead{$\lambda_{\rm rest}$} & \colhead{Flux} & \colhead{FWHM} & \colhead{v$_{\rm off}$}\\
\colhead{} & \colhead{($\mu$m)} & \colhead{(10$^{-17}$ erg cm$^{-2}$ s$^{-1}$)} & \colhead{(km s$^{-1}$)} & \colhead{(km s$^{-1}$)}}
\decimals
\startdata
\midrule
Hydrogen Recombination Lines & & & & \\
\midrule
Br $\eta$ 11-4 & 1.6806 & ${6.46}^{4.18}_{6.23}$ & ${371.4}^{1.4}_{1.4}$ & ${-0.05}^{0.53}_{0.56}$\\
Br $\zeta$ 10-4 & 1.7362 & ${8.71}^{4.60}_{5.16}$ & ${367.1}^{3.6}_{3.3}$ & ${-0.06}^{0.75}_{1.07}$\\
Br $\epsilon$ 9-4 & 1.8174 & ${11.57}^{4.67}_{5.74}$ & ${366.7}^{1.9}_{1.6}$ & ${0.05}^{0.57}_{0.57}$\\
Pa $\alpha$ 4-3 & 1.8751 & ${302.49}^{17.25}_{17.18}$ & ${378.0}^{1.2}_{1.2}$ & ${0.07}^{0.61}_{0.44}$\\
Br $\delta$ 8-4 & 1.9446 & ${16.78}^{5.29}_{5.69}$ & ${366.9}^{1.6}_{1.4}$ & ${0.07}^{0.72}_{0.79}$\\
Br $\gamma$ 7-4 & 2.1655 & ${27.58}^{1.75}_{2.02}$ & ${344.9}^{23.3}_{24.8}$ & ${5.44}^{8.34}_{10.85}$\\
Pf $\omicron$ 20-5 & 2.4307 & ${0.59}^{0.23}_{0.31}$ & ${315.5}^{58.0}_{49.8}$ & ${15.88}^{17.70}_{16.49}$\\
Pf $\xi$ 19-5 & 2.4483 & ${0.69}^{0.25}_{0.30}$ & ${276.2}^{69.1}_{55.5}$ & ${-12.98}^{24.15}_{25.60}$\\
Pf $\nu$ 18-5 & 2.4693 & ${0.78}^{0.31}_{0.28}$ & ${301.9}^{43.5}_{41.4}$ & ${22.29}^{26.91}_{26.05}$\\
Pf $\mu$ 17-5 & 2.4946 & ${1.13}^{0.35}_{0.41}$ & ${309.1}^{70.8}_{56.2}$ & ${21.15}^{21.17}_{20.07}$\\
Pf $\kappa$ 15-5 & 2.5636 & ${1.63}^{0.25}_{0.29}$ & ${310.0}^{37.8}_{34.7}$ & ${-1.26}^{21.20}_{18.97}$\\
Br $\beta$ 6-4 & 2.6252 & ${48.19}^{0.62}_{0.59}$ & ${307.2}^{3.1}_{3.1}$ & ${-9.85}^{1.46}_{1.51}$\\
Pf $\theta$ 13-5 & 2.6744 & ${2.40}^{0.33}_{0.33}$ & ${290.9}^{21.9}_{20.8}$ & ${-20.11}^{12.69}_{14.13}$\\
Pf $\eta$ 12-5 & 2.7575 & ${3.10}^{0.33}_{0.41}$ & ${256.3}^{32.8}_{29.1}$ & ${19.45}^{7.99}_{8.21}$\\
Pf $\zeta$ 11-5 & 2.8722 & ${3.73}^{0.36}_{0.42}$ & ${277.3}^{24.7}_{23.7}$ & ${24.31}^{8.65}_{8.03}$\\
Pf $\epsilon$ 10-5 & 3.0384 & ${5.32}^{1.09}_{1.11}$ & ${296.4}^{50.0}_{39.1}$ & ${-34.84}^{23.25}_{24.13}$\\
Pf $\delta$ 9-5 & 3.2961 & ${8.05}^{0.48}_{0.52}$ & ${410.2}^{28.5}_{24.0}$ & ${-3.75}^{8.41}_{9.01}$\\
Pf $\gamma$ 8-5 & 3.7395 & ${10.45}^{11.12}_{17.71}$ & ${466.9}^{185.6}_{331.0}$ & ${-15.74}^{101.06}_{112.87}$\\
Hu $\kappa$ 16-6 & 3.8184 & ${2.85}^{108.40}_{16.72}$ & ${229.9}^{443.7}_{229.9}$ & ${5.08}^{222.62}_{203.94}$\\
Hu $\iota$ 15-6 & 3.9065 & ${2.88}^{109.79}_{13.00}$ & ${253.8}^{399.7}_{253.8}$ & ${7.33}^{175.83}_{159.05}$\\
Hu $\theta$ 14-6 & 4.0198 & ${4.50}^{7.33}_{12.50}$ & ${370.5}^{161.7}_{190.6}$ & ${-10.46}^{130.42}_{155.98}$\\
Br $\alpha$ 5-4 & 4.0511 & ${84.52}^{21.82}_{25.56}$ & ${361.6}^{82.4}_{77.4}$ & ${-5.95}^{39.19}_{44.67}$\\
Hu $\eta$ 13-6 & 4.1697 & ${4.27}^{6.56}_{11.09}$ & ${344.8}^{186.7}_{208.9}$ & ${-9.57}^{117.76}_{100.86}$\\
Hu $\zeta$ 12-6 & 4.3753 & ${3.87}^{7.07}_{17.76}$ & ${338.5}^{304.9}_{202.6}$ & ${-5.60}^{65.02}_{68.15}$\\
Pf $\beta$ 7-5 & 4.6525 & ${16.61}^{5.55}_{4.74}$ & ${302.5}^{125.9}_{102.3}$ & ${-6.26}^{46.57}_{42.87}$\\
Hu $\epsilon$ 11-6 & 4.6712 & ${2.85}^{3.81}_{12.62}$ & ${354.7}^{335.4}_{218.8}$ & ${10.31}^{96.15}_{146.94}$\\
Hu $\delta$ 10-6 & 5.1273 & ${3.27}^{4.57}_{16.38}$ & ${287.8}^{364.6}_{152.0}$ & ${-7.88}^{106.25}_{125.47}$\\
\midrule
$H_{2}$ Lines & & & & \\
\midrule
7-5 Q(1) & 1.7283 & ${1.52}^{2.42}_{4.15}$ & ${367.8}^{2.6}_{2.6}$ & ${-0.01}^{0.97}_{1.04}$\\
4-1 O(20) & 1.7331 & ${2.26}^{3.00}_{5.43}$ & ${366.4}^{2.0}_{1.8}$ & ${-0.02}^{0.84}_{0.80}$\\
2-1 S(16) & 1.7461 & ${2.48}^{2.85}_{5.43}$ & ${367.0}^{2.3}_{2.4}$ & ${0.05}^{0.82}_{0.71}$\\
8-6 S(1) & 1.7635 & ${2.51}^{3.25}_{6.37}$ & ${368.7}^{1.9}_{1.8}$ & ${0.05}^{1.28}_{1.31}$\\
1-0 S(5) & 1.8353 & ${2.13}^{2.64}_{5.41}$ & ${366.9}^{1.8}_{1.5}$ & ${0.00}^{0.85}_{0.87}$\\
1-0 S(4) & 1.8914 & ${1.92}^{2.13}_{4.21}$ & ${367.4}^{1.5}_{1.3}$ & ${-0.07}^{0.88}_{1.08}$\\
9-7 S(1) & 1.9424 & ${2.08}^{2.35}_{4.84}$ & ${370.3}^{1.7}_{1.6}$ & ${0.04}^{0.95}_{0.92}$\\
1-0 S(3) & 1.9570 & ${3.72}^{3.14}_{5.06}$ & ${366.4}^{0.9}_{1.0}$ & ${-0.07}^{1.10}_{1.02}$\\
9-7 S(0) & 1.9868 & ${1.41}^{2.20}_{4.47}$ & ${378.6}^{0.9}_{1.5}$ & ${0.05}^{0.81}_{0.68}$\\
3-1 Q(23) & 2.0045 & ${1.17}^{1.45}_{3.22}$ & ${366.6}^{2.0}_{2.6}$ & ${-0.05}^{0.81}_{1.01}$\\
1-0 S(2) & 2.0332 & ${3.01}^{2.61}_{4.59}$ & ${367.3}^{2.7}_{2.4}$ & ${-0.07}^{0.86}_{0.80}$\\
8-6 O(3) & 2.0413 & ${1.66}^{2.50}_{4.32}$ & ${374.9}^{2.4}_{2.2}$ & ${-0.02}^{0.51}_{0.52}$\\
2-1 S(3) & 2.0729 & ${1.62}^{2.07}_{3.90}$ & ${366.6}^{2.5}_{2.5}$ & ${-0.03}^{0.56}_{0.58}$\\
1-0 S(1) & 2.1212 & ${5.20}^{0.96}_{1.01}$ & ${355.2}^{67.0}_{51.4}$ & ${-9.88}^{17.61}_{19.77}$\\
3-1 O(13) & 2.1534 & ${0.55}^{0.43}_{1.01}$ & ${213.8}^{55.4}_{77.7}$ & ${13.35}^{13.12}_{20.88}$\\
4-3 S(5) & 2.2004 & ${0.21}^{8.64}_{1.22}$ & ${136.3}^{115.1}_{136.3}$ & ${-51.56}^{42.43}_{30.56}$\\
1-0 S(0) & 2.2227 & ${1.39}^{0.42}_{0.63}$ & ${236.4}^{35.9}_{77.5}$ & ${6.20}^{8.28}_{8.38}$\\
2-1 S(1) & 2.2471 & ${1.88}^{0.72}_{0.96}$ & ${303.5}^{40.1}_{42.0}$ & ${6.92}^{23.70}_{29.70}$\\
3-2 S(2) & 2.2864 & ${0.29}^{0.31}_{0.73}$ & ${160.5}^{104.3}_{24.3}$ & ${28.64}^{16.27}_{18.64}$\\
2-1 S(0) & 2.3550 & ${1.24}^{0.30}_{0.32}$ & ${332.8}^{44.4}_{37.1}$ & ${-20.00}^{24.62}_{23.13}$\\
3-2 S(1) & 2.3858 & ${0.91}^{0.28}_{0.28}$ & ${281.2}^{76.4}_{66.7}$ & ${25.40}^{22.30}_{19.93}$\\
1-0 Q(1) & 2.4059 & ${6.57}^{0.33}_{0.35}$ & ${338.9}^{15.7}_{12.3}$ & ${25.03}^{7.95}_{7.47}$\\
1-0 Q(2) & 2.4128 & ${2.74}^{0.34}_{0.35}$ & ${324.5}^{45.6}_{43.1}$ & ${16.56}^{11.09}_{12.98}$\\
1-0 Q(3) & 2.4231 & ${3.70}^{0.32}_{0.35}$ & ${311.1}^{23.4}_{23.9}$ & ${0.74}^{11.56}_{11.97}$\\
1-0 Q(4) & 2.4368 & ${1.38}^{0.26}_{0.31}$ & ${331.8}^{35.7}_{36.0}$ & ${-21.66}^{19.77}_{20.50}$\\
1-0 Q(5) & 2.4541 & ${1.72}^{0.30}_{0.33}$ & ${306.7}^{30.6}_{31.7}$ & ${-41.48}^{18.79}_{19.30}$\\
1-0 Q(6) & 2.4749 & ${0.48}^{0.27}_{0.38}$ & ${337.0}^{44.3}_{42.8}$ & ${-117.61}^{23.65}_{19.71}$\\
5-4 S(21) & 2.4994 & ${1.24}^{0.36}_{0.45}$ & ${405.9}^{66.9}_{60.9}$ & ${18.23}^{21.95}_{20.64}$\\
1-0 Q(8) & 2.5274 & ${1.21}^{0.21}_{0.24}$ & ${288.7}^{37.4}_{36.7}$ & ${-119.73}^{19.88}_{20.59}$\\
4-3 S(1) & 2.5407 & ${0.59}^{0.23}_{0.29}$ & ${384.3}^{120.5}_{117.3}$ & ${-32.62}^{58.40}_{53.11}$\\
2-1 Q(1) & 2.5503 & ${2.65}^{0.27}_{0.26}$ & ${269.4}^{34.2}_{34.4}$ & ${-27.72}^{10.82}_{11.54}$\\
2-1 Q(2) & 2.5578 & ${2.03}^{0.35}_{0.36}$ & ${419.2}^{76.0}_{67.7}$ & ${1.69}^{27.85}_{25.67}$\\
2-1 Q(3) & 2.5691 & ${1.73}^{0.31}_{0.34}$ & ${306.9}^{57.7}_{55.3}$ & ${-19.69}^{19.12}_{19.28}$\\
2-1 Q(4) & 2.5843 & ${0.99}^{0.39}_{0.47}$ & ${657.5}^{63.7}_{165.2}$ & ${-92.72}^{90.66}_{78.29}$\\
2-1 Q(5) & 2.6033 & ${0.45}^{0.26}_{0.33}$ & ${251.4}^{127.6}_{115.3}$ & ${24.65}^{47.48}_{60.42}$\\
10-8 Q(8) & 2.6115 & ${1.90}^{0.31}_{0.36}$ & ${306.9}^{38.5}_{30.3}$ & ${33.65}^{20.03}_{20.15}$\\
2-1 Q(7) & 2.6531 & ${0.28}^{0.22}_{0.29}$ & ${208.3}^{186.8}_{72.1}$ & ${60.62}^{61.74}_{77.31}$\\
3-2 Q(1) & 2.7095 & ${0.83}^{0.21}_{0.22}$ & ${249.2}^{23.0}_{53.7}$ & ${-3.80}^{3.71}_{4.14}$\\
9-7 Q(13) & 2.7171 & ${0.77}^{0.22}_{0.25}$ & ${256.4}^{15.9}_{43.4}$ & ${-23.96}^{6.12}_{4.22}$\\
7-5 Q(19) & 2.7264 & ${0.00}^{0.00}_{0.00}$ & ${0.0}^{136.1}_{0.0}$ & ${0.08}^{12.31}_{16.05}$\\
3-2 Q(3) & 2.7305 & ${0.70}^{0.39}_{0.61}$ & ${262.3}^{27.2}_{23.5}$ & ${-8.77}^{11.27}_{10.12}$\\
3-2 Q(4) & 2.7473 & ${0.29}^{0.22}_{0.42}$ & ${185.3}^{39.1}_{24.1}$ & ${-52.90}^{16.88}_{17.06}$\\
2-1 O(2) & 2.7854 & ${1.04}^{0.30}_{0.35}$ & ${238.3}^{32.8}_{32.3}$ & ${9.14}^{13.35}_{13.38}$\\
1-0 O(3) & 2.8018 & ${5.61}^{0.49}_{0.46}$ & ${232.7}^{26.0}_{25.6}$ & ${-13.11}^{7.98}_{7.14}$\\
4-3 Q(1) & 2.8874 & ${0.65}^{0.28}_{0.36}$ & ${253.6}^{31.7}_{37.1}$ & ${-5.45}^{11.51}_{14.60}$\\
4-3 Q(3) & 2.9111 & ${0.15}^{0.14}_{0.88}$ & ${136.1}^{0.0}_{0.0}$ & ${14.57}^{2.16}_{2.76}$\\
3-2 O(2) & 2.9613 & ${0.69}^{0.86}_{1.70}$ & ${328.3}^{333.3}_{192.2}$ & ${13.21}^{131.32}_{143.21}$\\
2-1 O(3) & 2.9733 & ${1.80}^{0.87}_{1.20}$ & ${283.6}^{164.4}_{147.5}$ & ${-13.61}^{87.38}_{64.27}$\\
1-0 O(4) & 3.0030 & ${2.08}^{0.85}_{1.21}$ & ${261.5}^{182.3}_{125.4}$ & ${-24.70}^{61.63}_{60.30}$\\
3-2 O(3) & 3.1629 & ${1.25}^{0.43}_{0.45}$ & ${437.0}^{37.5}_{37.1}$ & ${-38.98}^{13.07}_{14.06}$\\
2-1 O(4) & 3.1889 & ${0.80}^{0.36}_{0.42}$ & ${286.8}^{38.2}_{38.6}$ & ${-20.89}^{15.52}_{12.76}$\\
10-7 O(16) & 3.2060 & ${2.26}^{0.34}_{0.50}$ & ${332.7}^{27.0}_{24.2}$ & ${14.15}^{18.90}_{15.40}$\\
7-5 O(14) & 3.2346 & ${1.37}^{0.37}_{0.50}$ & ${313.7}^{45.8}_{39.5}$ & ${-32.66}^{15.60}_{17.78}$\\
4-3 O(3) & 3.3755 & ${0.16}^{0.19}_{0.45}$ & ${191.0}^{77.4}_{55.1}$ & ${42.86}^{15.31}_{16.80}$\\
3-2 O(4) & 3.3949 & ${0.97}^{0.37}_{0.53}$ & ${399.4}^{28.2}_{35.4}$ & ${-12.67}^{21.71}_{17.91}$\\
2-1 O(5) & 3.4369 & ${0.53}^{0.22}_{0.29}$ & ${221.0}^{50.7}_{64.2}$ & ${42.74}^{7.71}_{8.15}$\\
1-0 O(6) & 3.4998 & ${0.29}^{0.28}_{1.11}$ & ${189.9}^{78.2}_{54.0}$ & ${-27.76}^{10.23}_{10.07}$\\
3-2 Q(18) & 3.5228 & ${0.51}^{0.31}_{0.45}$ & ${232.3}^{39.5}_{83.4}$ & ${-109.35}^{10.36}_{9.56}$\\
0-0 S(11) & 4.1799 & ${3.76}^{6.67}_{9.29}$ & ${369.7}^{176.9}_{154.3}$ & ${-4.34}^{78.55}_{68.53}$\\
0-0 S(9) & 4.6933 & ${2.51}^{3.09}_{6.90}$ & ${376.9}^{257.1}_{241.0}$ & ${15.90}^{159.17}_{156.92}$\\
\midrule
He I Lines & & & & \\
\midrule
 & 1.7003 & ${4.31}^{3.21}_{8.06}$ & ${366.7}^{1.8}_{2.2}$ & ${-0.02}^{1.18}_{0.86}$\\
 & 1.8685 & ${16.35}^{5.48}_{6.07}$ & ${366.7}^{2.5}_{2.5}$ & ${-0.01}^{0.79}_{0.74}$\\
 & 1.9089 & ${1.36}^{1.81}_{3.67}$ & ${372.4}^{2.6}_{2.2}$ & ${0.03}^{1.14}_{1.13}$\\
 & 1.9543 & ${5.10}^{4.94}_{9.35}$ & ${366.3}^{1.8}_{2.0}$ & ${0.03}^{0.76}_{0.84}$\\
 & 2.0580 & ${11.23}^{3.69}_{5.19}$ & ${371.1}^{2.1}_{2.0}$ & ${0.07}^{0.88}_{1.16}$\\
 & 2.1120 & ${2.10}^{0.83}_{0.93}$ & ${396.7}^{36.0}_{38.5}$ & ${-10.85}^{25.27}_{16.87}$\\
 & 2.6185 & ${1.14}^{0.41}_{0.43}$ & ${476.4}^{139.7}_{139.4}$ & ${55.46}^{66.32}_{70.38}$\\
 & 4.0366 & ${4.56}^{6.98}_{13.05}$ & ${360.6}^{160.2}_{224.7}$ & ${-5.02}^{60.81}_{57.12}$\\
 & 4.2947 & ${11.04}^{11.68}_{19.85}$ & ${377.4}^{184.6}_{126.1}$ & ${-1.12}^{60.41}_{64.61}$\\
\enddata
\label{tab:ident}
\end{deluxetable*}